# Quantum AI simulator using a hybrid CPU–FPGA approach

Teppei Suzuki[1], Tsubasa Miyazaki[1], Toshiki Inaritai[1] & Takahiro Otsuka[1]

[1] *Research and Development Center, SCSK Corporation, Toyosu Front, 3-2-20 Toyosu, Koto-ku, Tokyo 135-8110, Japan*

`E-mail: tep.suzuki@scsk.jp`

**Abstract:** The quantum kernel method has attracted considerable attention in the field of quantum machine learning. However, exploring the applicability of quantum kernels in more realistic settings has been hindered by the number of physical qubits current noisy quantum computers have, thereby limiting the number of features encoded for quantum kernels. Hence, there is a need for an efficient, application-specific simulator for quantum computing by using classical technology. Here we focus on quantum kernels empirically designed for image classification and demonstrate a field programmable gate arrays (FPGA) implementation. We show that the quantum kernel estimation by our heterogeneous CPU–FPGA computing is 470 times faster than that by a conventional CPU implementation. The co-design of our application-specific quantum kernel and its efficient FPGA implementation enabled us to perform one of the largest numerical simulations of a gate-based quantum kernel in terms of features, up to 780-dimensional features. We apply our quantum kernel to classification tasks using the Fashion-MNIST dataset and show that our quantum kernel is comparable to Gaussian kernels with the optimized hyperparameter.

Introduction

Quantum computing [1] is an emerging technology that could transform many areas of industries and scientific research, including finance [2], chemistry [3], and machine learning (ML) [4, 5]. In particular, quantum machine learning (QML) [4, 5, 6–17] has received considerable attention at a rapid rate, indicating that QML is a plausible candidate for the practical application of near-term quantum devices. While early fault-tolerant quantum computing has been demonstrated recently [18], noisy intermediate-scale quantum (NISQ) processors are currently available through various hardware platforms with ~10–100 physical qubits. However, the number of physical qubits today's NISQ computers have is generally insufficient to explore practical applications of QML. Therefore, there is a need for application-specific quantum computing simulators to explore and validate the practicality of QML in real-world settings.

The quantum kernel method is a NISQ algorithm in the framework of the hybrid quantum–classical approach [19, 20] and can also be feasible on current NISQ computers with shallow quantum circuits [9, 12, 13, 16, 17]. In the quantum kernel method, a quantum feature map can be described explicitly by a quantum circuit and the quantum kernel entry can be estimated by measuring the inner product of the quantum feature map [8, 9]. The calculation of quantum kernels when using real devices or general-purpose simulators based on quantum assembly language (QASM) requires a number of measurements to obtain the quantum kernel entries (note that measurements are a key part of the QASM simulator, which handles measurements by collapsing the state of the qubit according to the probabilities predicted by quantum mechanics). Commonly used quantum kernels inspired by instantaneous quantum polynomials (IQP) [9] can be computationally prohibitive on classical computers as the number of qubits increases; for instance, the number of entangled qubits in the simulation of quantum kernels using state-of-the-art classical platforms is 30 [11]. On the other hand, it becomes challenging to reliably estimate such quantum kernels using near-term quantum devices with increasing size in circuits, owing to expensive gate costs, low gate fidelities, and different qubit connectivities. The above points can be a drawback in exploring practical applications of quantum kernels since machine learning models typically improve performance by increasing training data samples or expanding the number of input features. There is still a gap between theoretical developments and practical applications in the quantum kernel method.

To bridge the gap between theory and practice in the quantum kernel method, in this paper, we focus on an application-specific quantum kernel that can be applied to image data with a large number of features. To this end, we demonstrate an implementation of an efficient quantum AI simulator by using a heterogeneous classical computing platform. Our approach is highly customized for our specific tasks at the hardware level and the main objective of our simulator differs from a general-purpose quantum simulator, which is designed to be versatile and to perform a range of quantum algorithms. Until now, there have been considerable efforts to develop quantum computing simulators [21–27]. Among hardware implementations, field programmable gate arrays (FPGA) are one of the desirable platforms, because FPGA has the properties of efficient parallelism, low latency, and customization. FPGAs comprise programmable logic blocks that can be interconnected to perform parallel processing, allowing each logic block to perform a specific task simultaneously. FPGAs can also be customized to perform specific tasks using hardware description languages such as Verilog. Herein, we co-design application-specific quantum kernels and our FPGA architecture, which allows efficient numerical simulations. FPGA has been successfully applied to fault-tolerant quantum algorithms such as Grover's algorithm [28–30], quantum Fourier transform [28–31], and Deutsch's algorithm [32]. However, an FPGA implementation of quantum kernels has been unexplored and the present study is the first demonstration of a gate-based quantum kernel simulator using an FPGA platform.

The rest of the paper is organized as follows. We provide a brief introduction to support vector machine (SVM) and describe our quantum feature map that is useful for image classification. Then we explain the overview of our quantum AI simulator using a heterogeneous CPU–FPGA computing. From an algorithmic point of view, the quantum kernel method can be divided into the quantum kernel estimation and the rest of the tasks. The simulation of the quantum kernel can be computationally demanding; hence, the workload can be accelerated by FPGA hardware. On the other hand, the rest of the tasks, such as dimensionality reduction and the optimization of machine learning parameters, can be efficiently performed on the CPU using existing classical libraries. The FPGA implementation of the quantum kernel is checked in terms of both numerical precision and hardware acceleration. We apply our quantum kernel simulator to binary and multiclass classification for a range of input features using the Fashion-MNIST dataset. Then we summarize our conclusions.

Results

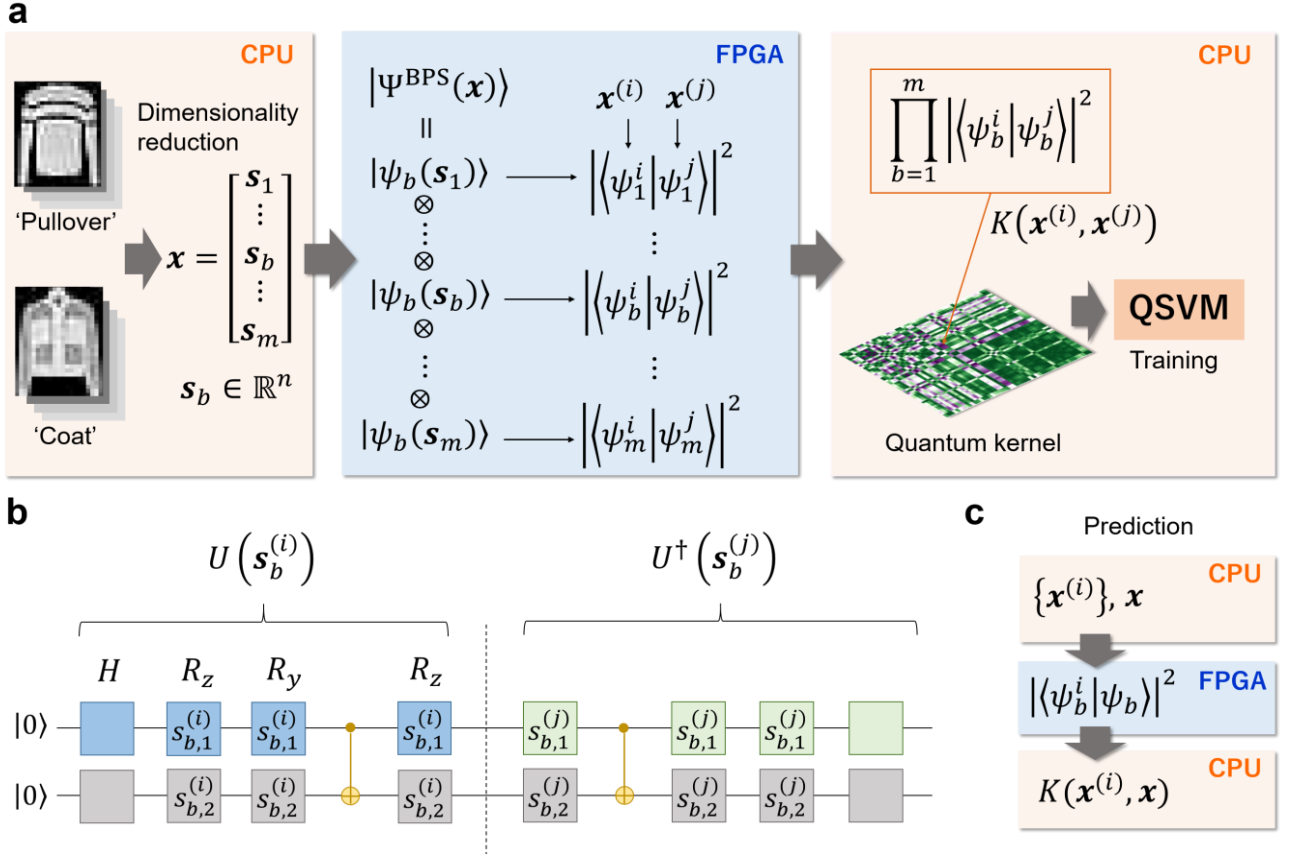

**Figure 1.** Schematic representation of our quantum AI simulator using a hybrid CPU–FPGA approach. (**a**) PCA is used to reduce the dimension of the original data from Fashion-MNIST; a range of features from $d = 4$ to $d = 780$ can be used for machine learning. Then PCA-reduced features are sent from the CPU to the FPGA. We calculate the square of the norm of the inner products for each block wavefunction $|\psi_b\rangle$ of the quantum feature map. This process is repeated for all the pairs of the data points (i.e., $N^2/2$ times). The data are then sent back to the CPU. A kernel matrix value can be obtained by multiplying $m$ blocks. After all the quantum kernel entries are computed, the SVM algorithm is performed on the CPU. (**b**) 2-qubit example of a quantum circuit that performs the estimation of the quantum kernel element. For the entangling gate, the CNOT gate is used. The quantum circuit is simulated on FPGA using the procedure described in the text (see Methods). (**c**) Test process: the decision function can be computed using the hybrid CPU–FPGA scheme. Details of computational resources in the cloud system (FPGA and CPU) are given in Methods.

**Quantum support vector machine.** The quantum kernel method is one of the most important algorithms in QML techniques and many studies have been reported [4, 5, 8, 9, 11–17]. In the classical kernel method [33, 34] the inner product of the feature map is represented by kernel functions, which implicitly use the Hilbert space; on the other hand, the quantum kernel explicitly defines a quantum feature map by means of a quantum state $|\phi(x)\rangle$ for $d$-dimensional input vectors $x \in \mathbb{R}^d$. The quantum kernel matrix $K(x, x')$ can be estimated by calculating the inner product of the quantum feature map [8, 9]:

$$K(x, x') = |\langle \phi(x)|\phi(x')\rangle|^2. \tag{1}$$

For binary classification in the framework of SVM, one can obtain a support vector classifier that estimates the label for a new datum $x$:

$$y = \text{sgn}\left(\sum_i y_i \alpha_i^* K(x^{(i)}, x) + b^*\right), \tag{2}$$

where $y_i \in \{+1, -1\}$ and parameters $\{\alpha_i^*\}$ and $b^*$ are the optimal parameters obtained in the training phase [34]. In the hybrid quantum–classical algorithm, the training phase can be performed on classical computers, whereas the quantum kernel entries can be computed by NISQ computers or quantum computing simulators; such methodology is called the quantum SVM (QSVM). The NISQ computation of the quantum kernel requires many quantum measurements to obtain a quantum kernel entry with statistically reliable accuracy. For example, a value for each computational-basis measurement is zero or one. For each quantum kernel entry, $O(N^2)$ shots are required with respect to the number of data samples $N$, resulting in the computational complexity of $O(N^4/\varepsilon^2)$ operations with the maximum error $\varepsilon$, in order to obtain all the quantum kernel entries [9]. Such computational complexity prohibits us from developing and validating quantum kernels as the number of data samples grows. Also, the number of entangling qubits with different connectivities in the previously proposed quantum kernels is increased with qubit count [9], which requires additional computational resources.

To address these issues, here we introduce a shallow, fixed-depth quantum circuit that can be applied to a quantum kernel for a larger number of input features. In the previously proposed quantum kernels based on IQP circuits [9], the number of dimensional features is typically set to the number of entangled qubits [9, 11, 14, 15]. IQP circuits are a subclass of quantum circuits that cannot be classically efficiently simulated unless the polynomial-time hierarchy collapses to the third level [35]. Here, an IQP circuit is a circuit where a Hadamard gate is applied to each qubit at the beginning and end of the computation, but the rest of the gates are diagonal. In the context of the quantum kernel method, researchers have typically used a more specific type of IQP, called the *ZZ* feature map [9]. In the *ZZ* feature map, the connectivity of qubits is achieved in a pair-wise manner, resulting in $n(n-1)/2$ interactions, where $n$ is the number of qubits. This leads to a rapid expansion of expressibility and results in a deterioration of generalization performance as qubit count increases [11, 14, 15]. Our approach aims to simplify the quantum feature map, limit the extent to which qubits are entangled, and control the capacity of our QML model, while increasing the number of input features. This framework can handle several hundreds of input features in QSVM. For the $mn$-dimensional input vector $\mathbf{x} = [\mathbf{s}_1, \mathbf{s}_2, \cdots, \mathbf{s}_m]^{\mathrm{T}} \in \mathbb{R}^{mn}$, where $\mathbf{s}_b$ is the $n$-dimensional vector $\mathbf{s}_b = [s_{b,1}, s_{b,2}, \cdots, s_{b,n}]^{\mathrm{T}}$, we consider a block product state (BPS) wavefunction [36]:

$$|\Psi^{\mathrm{BPS}}(\mathbf{x})\rangle = |\psi_1(\mathbf{s}_1)\rangle \otimes |\psi_2(\mathbf{s}_2)\rangle \otimes \cdots \otimes |\psi_m(\mathbf{s}_m)\rangle, \tag{3}$$

where

$$|\psi_b(\mathbf{s}_b)\rangle = \left(\otimes_{q=1}^{n} R_z(s_{b,q})\right) U_{2^n}^{\mathrm{ent}} \left(\otimes_{q=1}^{n} \left(R_y(s_{b,q}) R_z(s_{b,q}) H\right)\right) |0^{\otimes n}\rangle, \tag{4}$$

and

$$U_{2^n}^{\mathrm{ent}} := \prod_{q=1}^{n-1} \mathbf{CNOT}_{q,q+1}. \tag{5}$$

In the BPS wavefunction, a modest number of qubits can be entangled within each block (in our numerical simulations, $n$ was varied from 2, 3, and 6); and for the wavefunction $|\psi_b(\mathbf{s}_b)\rangle$, each component $s_{b,q}$ is encoded three times as the input angle for the ration operator gates (i.e., $s_{b,q}$ is encoded in the $R_z$ gate, in the $R_y$ gate, and again in the $R_z$ gate in Eq. (4)). Such kind of redundant encoding leads to the better performance of QML models based on angle encoding [37]. The state $|\psi_b(\mathbf{s}_b)\rangle$ is related to matrix product states, a class of tensor networks that have been used for the study of ground states of quantum systems and recently for machine learning. The connectivity of qubits in Eq. (4) is limited to their nearest neighbors, resulting in $(n-1)$ interactions. The idea of BPS has been originally used for ML models based on tensor networks [36]; yet, to our knowledge, this kind of BPS-based quantum feature map has not been applied to quantum kernels. In this work, we will show that such a feature map can be used for QSVM. The kernel associated with the quantum feature map defined by Eq. (3) can be given by

$$K(\mathbf{x}^{(i)}, \mathbf{x}^{(j)}) = |\langle \Psi^{\mathrm{BPS}}(\mathbf{x}^{(i)})|\Psi^{\mathrm{BPS}}(\mathbf{x}^{(j)})\rangle|^2 = \prod_{b=1}^{m} \left|\left\langle \psi_b\left(\mathbf{s}_b^{(i)}\right) \middle| \psi_b\left(\mathbf{s}_b^{(j)}\right)\right\rangle\right|^2. \tag{6}$$

The number of blocks $m$ can be varied in order to allow a larger number of input features depending on different datasets. Another interesting aspect is that the quantum kernel is not translation invariant, which means that the quantum kernel does not depend solely on the distance of input vectors, in contrast with Gaussian kernels. A computational benefit of our approach is that the calculation of the quantum kernel can be divided into $m$ computational tasks, allowing an efficient computation on classical computers. In particular, $|\langle\psi_b^i|\psi_b^j\rangle|^2$ in Eq. (6) can be computed separately; hence, each task can now be efficiently simulated through FPGA acceleration and the multiplication can then be performed on CPU.

Quantum AI simulator using a hybrid CPU–FPGA approach. By co-designing FPGA architecture and a quantum kernel given by a shallow quantum circuit, we implemented a fast and efficient quantum AI simulator using a heterogeneous computing approach (Fig. 1a). Details of computational resources in the cloud system (FPGA and CPU) are given in Methods. To begin with, using the principal component analysis (PCA) method [38] we conducted the dimensionality reduction of the $28 \times 28$ image data from the Fashion-MNIST dataset [39]; then the number of input features can be varied from $d = 4$ to $d = 780$. After obtaining PCA-reduced input vectors $x^{(i)} \in \mathbb{R}^d$, the input data are sent from CPU to the internal memory of an FPGA hardware via PCI express. Then, for each block wavefunction $|\psi_b(s_b)\rangle$ $(b = 1, \cdots, m)$ of the quantum feature map, we calculate the square of the norm of the inner products $|\langle\psi_b^i|\psi_b^j\rangle|^2$ (which is depicted in Fig. 1b) on our FPGA architecture in the following procedure: First, the sine and cosine of the input angles for quantum gates are computed using the COordinate Rotational DIgital Computer (CORDIC) algorithm [40]. Second, the square of the norm of the inner product can be calculated using the unitary matrices in Eq. (4), together with an efficient implementation of $n$-qubit entanglement. (The procedure is described in great detail in Methods and Supplementary Notes 1 and 2.) This process can be repeated for all the pairs of data samples, namely, for $N^2/2$ cycles. The processed, real-valued data are sent back to the CPU. The kernel matrix element will thus be calculated by the multiplication of $m$ blocks. After all the kernel entries are obtained, the training phase of the SVM can be performed on the CPU platform. In the test process (Fig. 1c), prediction can be done using the same FPGA acceleration with $\mathcal{O}(ND)$ operations, where $D$ is the number of test data.

FPGA implementation: numerical precision and acceleration. Herein we validate our FPGA implementation in terms of numerical precision and acceleration. We begin by comparing the quantum kernel values obtained by the FPGA platform and those obtained by the CPU platform (Fig. 2a–c). The norms of inner products $|\langle\psi_b^i|\psi_b^j\rangle|^2$ have values between 0 and 1. Such property along with a shallow circuit depth is amenable to the use of 16-bit fixed-point arithmetic in our FPGA architecture, which in turn makes the calculation faster with efficient hardware utilization. We also employed 64-bit floating-point arithmetic in the CPU platform to validate our FPGA implementation. The parity plot suggests the success of our FPGA implementation of the quantum kernel (Fig. 2c). The numerical deviation between the two hardware platforms was $\pm \sim 0.095\%$, indicating that there was a negligible loss of numerical accuracy.

Next, we compare the execution time for computing a kernel matrix (in the case of 6 entangled qubits) using the FPGA platform with the one obtained by our CPU implementation, as well as the one obtained by Qiskit Aer [21], a QASM quantum computing simulator (Fig. 2d). Measurement is a vital aspect of the simulation process in the QASM simulator, which handles measurements by collapsing the state of the qubit according to the probabilities determined by the state of the qubit. Therefore, in the QASM simulator, a number of shots are required to obtain the expectation value. In our CPU implementation, the kernel matrix entry is obtained directly by calculating the inner product of the state vectors. In particular, we used NumPy [41], which is a popular library for scientific computing and data analysis (note that the core of NumPy is implemented in C Language). For our particular tasks (in the case of 6 entangled qubits), the execution time by our CPU implementation is likely to be somewhat faster than that by the state-vector simulator; this is because the state-vector simulator tracks the quantum state of the system as it evolves through the circuit, resulting in a slowing down of the execution time. Thus, the plot for the state-vector simulation would be the upper side of the plot denoted by orange in Fig. 2d.

In our FPGA architecture, once the data are sent to the FPGA architecture, we used only the internal memory of the FPGA hardware without accessing the external (off-chip) memory, which circumvents the associated communication overhead (for more details of our FPGA architecture, see Supplementary Note 2). In addition, two more factors are responsible for the FPGA acceleration. First, an FPGA allows each programable logic block to perform a specific task simultaneously in an efficient manner. Second, an FPGA can be customized to perform specific tasks using the hardware description language, resulting in faster performance in comparison with CPUs.

In our FPGA implementation, all the kernel entries were computed in 4.1 ms at $N = 1000$; and the execution time including CPU–FPGA communication overhead was 15.4 ms at $N = 1000$. In other words, our FPGA implementation achieved a 1784 × improvement in comparison with the CPU counterpart. Also, the execution time including the communication overhead was 472 times faster (Fig. 2d); moreover, in comparison with the execution by a QASM simulator (assuming that the computation cost grows as $\mathcal{O}(N^4/\varepsilon^2)$ operations), a 10 million times speedup was accomplished at $N = 400$ (Fig. 2d). The results show that our FPGA implementation is highly efficient in terms of the number of data samples, with a modest number of entangling qubits (up to 6 qubits) being used in our quantum feature map. Owing to the limitation of the internal memory and digital single processors within an FPGA, however, our implementation technique will be prohibitive for $n$ large than 8. Nonetheless, for our machine-learning tasks, this can be overcome by dividing input features into a number of blocks, and each block's quantum kernel can be efficiently computed in FPGA. Thus, the FPGA-based simulator accelerates the numerical simulations of QSVM using our quantum kernel and allows us to validate its applicability to much larger features in quantum kernel methods.

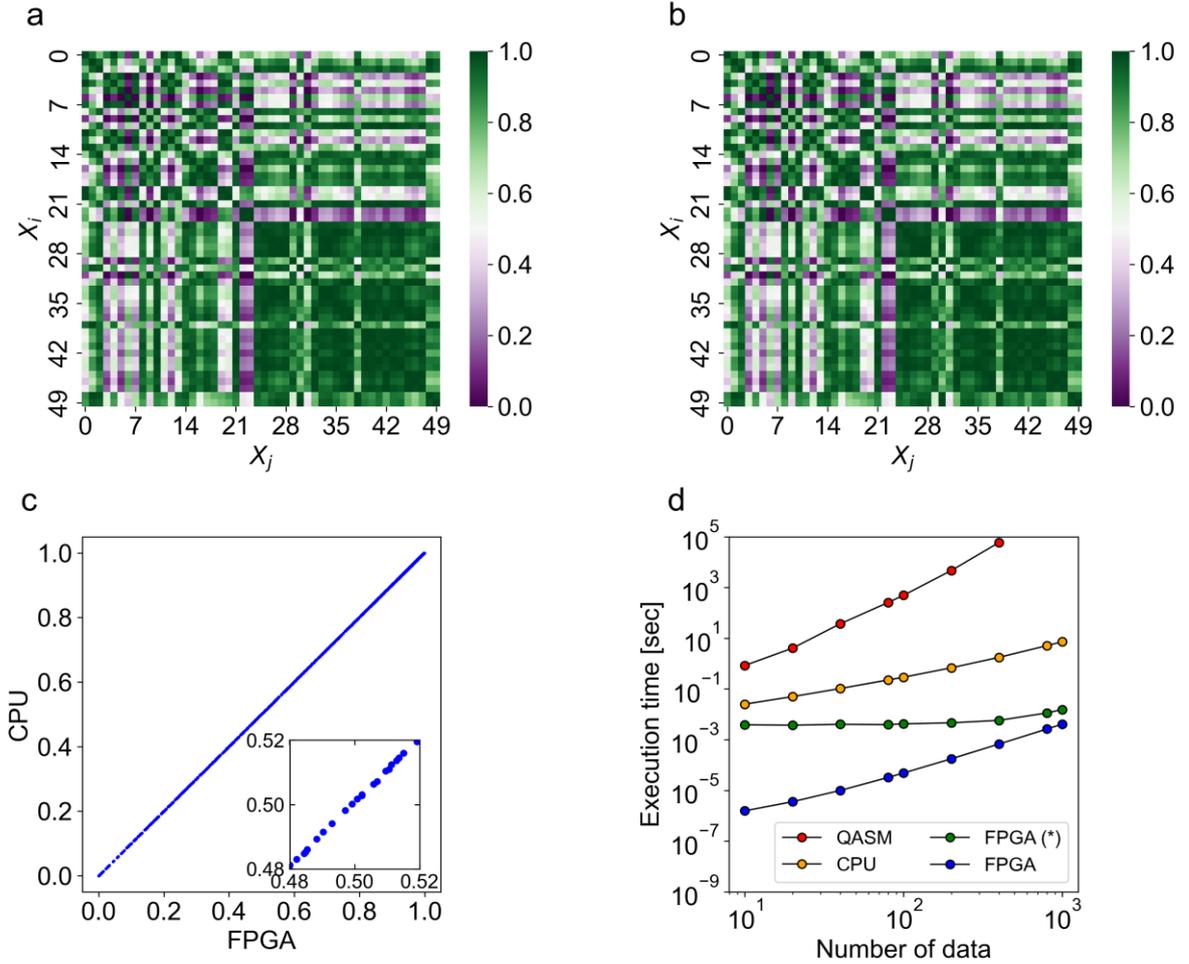

**Figure 2.** FPGA implementation of the quantum kernel and its execution time. The numerical simulations were performed on a 6-qubit quantum circuit that estimates the quantum kernel element. (**a**) Quantum kernel matrix obtained by an FPGA platform (16-bit fixed-point arithmetic). (**b**) Quantum kernel matrix obtained by a CPU platform (64-bit floating-point arithmetic). (**c**) Parity plot for the quantum kernel values obtained by CPU and FPGA platforms. Inset shows small differences between the two: the error between the two hardware platforms was $\pm$ ~0.095%. (**d**) Execution time with respect to the number of data $N$ for different platforms: FPGA, blue; FPGA (including CPU–FPGA communication overhead; denoted by the asterisk), green; CPU, orange; QASM quantum simulator (Qiskit Aer), red. Note that the FPGA execution including communication overhead was 472 times faster than that for the CPU counterpart at $N = 1000$.

**Binary classification on Fashion-MNIST dataset.** Having shown the accuracy and efficiency of our FPGA-based quantum kernel estimation, we now turn to the performance of our quantum kernel. To begin with, we trained classical and quantum SVMs on Fashion-MNIST and obtained 45 binary classifiers. Among 45 pairs of binary classification tasks from 10 categories of Fashion-MNIST [39] (0, t-shirt/top; 1, trouser; 2, pullover; 3, dress; 4, coat; 5, sandals; 6, shirt; 7, sneaker; 8, bag; 9, ankle boots), about half the pairs of classification tasks were relatively easy to distinguish. On the other hand, more challenging tasks such as pullover vs. shirt (2 vs. 6), pullover vs. coat (2 vs. 4), and coat vs. shirt (4 vs 6) classification tasks were somewhat difficult to distinguish (e.g., the images of pullovers are more similar to those of coats than to those of trousers). Hence, we focused on the three binary classification tasks and investigate the performance in detail (Fig. 3).

The performance of our quantum kernel without introducing any hyperparameter was comparable to that of the Gaussian kernel $\exp\left(-\gamma\|x^{(i)} - x^{(j)}\|^2\right)$ with the optimized bandwidth $\gamma$, for dimensions smaller than ~300 (Fig. 3a). Here, a key hyperparameter in the Gaussian kernel is the kernel bandwidth $\gamma$, which is known to affect the performance of kernel-based methods such as SVMs and is routinely optimized when SVMs are used in practice. The hyperparameter $\gamma$ controls the smoothness of the decision boundary in the SVM. Analogously, we introduced a scaling hyperparameter $\lambda$ (i.e., $x^{(i)} \leftarrow \lambda x^{(i)}$ in the quantum circuit) to improve the performance of QSVM. The role of $\lambda$ appears to be similar to the classical counterpart. The hyperparameter $\lambda$ can calibrate the angles of the rotation gates and directly affect the quantum feature map in the Hilbert space. From a physical point of view, changing the hyperparameter $\lambda$ in the quantum kernel is related to changing the total evolution time in the Hamiltonian evolution [15]. The best test accuracy for the quantum kernel was 0.87 at $d = 180, 190, 200$; whereas that for the classical kernel with the optimal bandwidth was 0.88 at $d = 190$. We found that introducing the scaling parameter $\lambda$ improved the performance of our quantum kernel for larger dimensions ($d > $ ~300), maintaining its comparable performance to the classical kernel, which is indicated by the blue dotted line in Fig. 3a (for the grid search over the hyperparameters of the classical and quantum kernels, see Supplementary Note 3).

The test accuracy obtained by our quantum kernel was improved by increasing the number of data samples $N$ (Fig. 3b). In particular, as the number $N$ was increased, the test accuracies for higher dimensional vectors tended to improve gradually (Fig. 3b). But the relatively sharp drop for dimensions higher than ~300 was difficult to overcome just by increasing $N$; nonetheless, the dimension $d$ that gave the best test accuracy was typically in the range between 100 and 200 for this particular application. We note that the drop in the test accuracy for higher dimensions can be overcome by optimizing the aforementioned scaling parameter (which will be discussed in multiclass classification).

The performance of our hyperparameter-free quantum kernel was competitive with the Gaussian kernel with the optimized bandwidth at $N > 1500$ (Fig. 3c), which might be beneficial for practical applications. The best test accuracies at $N = 2000$ and $N = 3000$ were 0.89 and 0.90, respectively, for both of the two kernels. For smaller numbers of data samples ($N < 1000$), the performance of our quantum kernel was slightly lower than the best classical counterpart. To understand the role of quantum entanglement, we investigated the effects of enlarging the number of entangled qubits. Increasing the number of entangled qubits (from 2 to 6 qubits per block) did not significantly change the performance for PCA-reduced input vectors (Fig. 3d); this kind of insensitiveness to quantum entanglement has been previously reported in an ML model based on tensor networks using BPS [36]. Our results probably indicate that the capacity of our quantum feature map is already sufficiently high even in the case of $n = 2$. However, this may not necessarily mean that quantum entanglement is unimportant; the CNOT entangling gate can make the quantum feature map more complex in comparison with no quantum entanglement.

Overall, the behavior of our quantum kernel is quite different from the previously used quantum kernels [9, 11–15]. The results suggest that our quantum kernel is comparable to the best classical kernel with good generalization performance for a range of features.

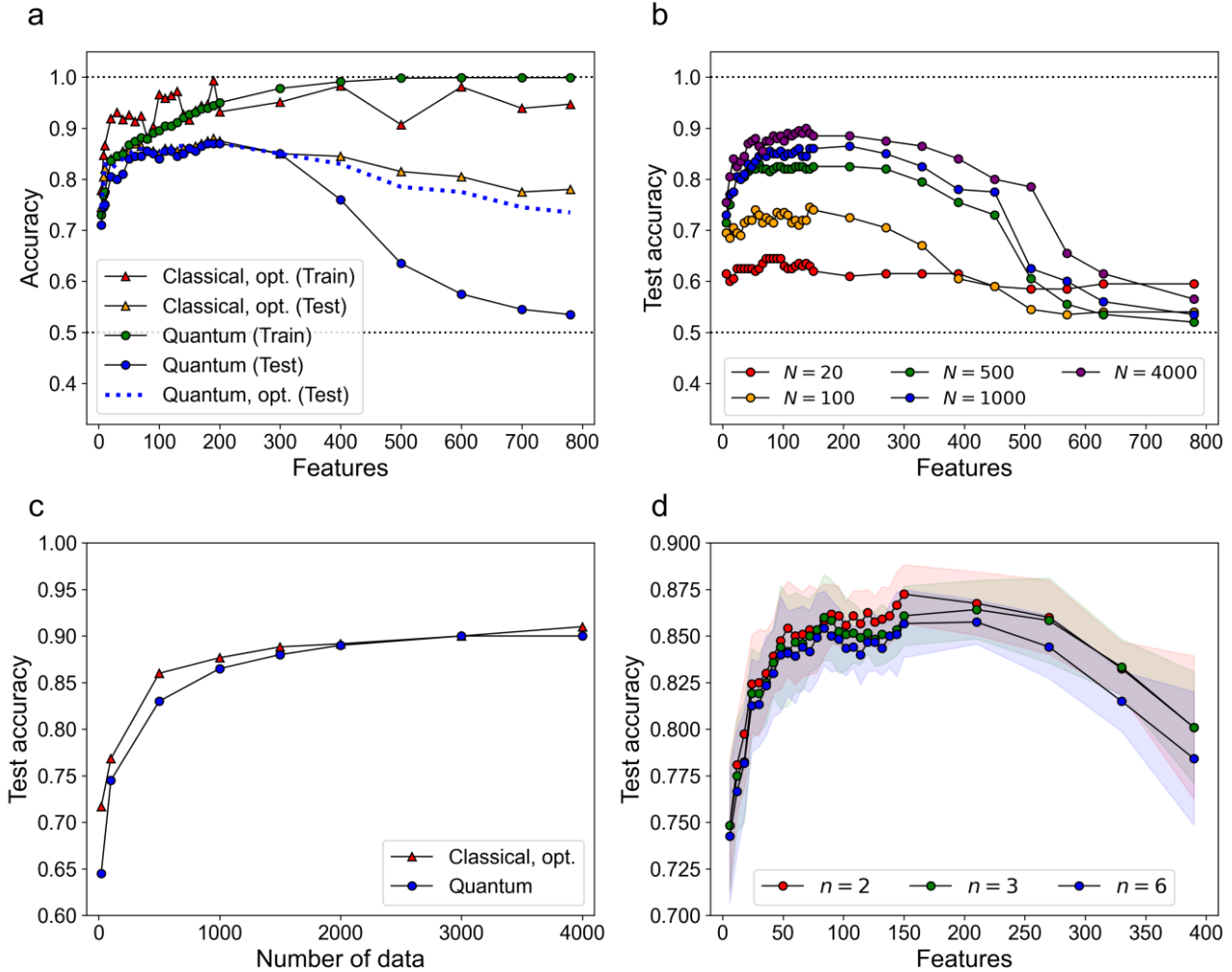

**Figure 3.** Train and test accuracies of the quantum kernel on the Fashion-MNIST dataset. (**a**) Training and test accuracies with a range of features from $d = 4$ to $d = 780$ using 1000 data samples. The coat vs. shirt (4 vs 6) classification task was used. The performance of the Gaussian kernel with the optimized bandwidth for each dimension (train: red triangle; test: yellow triangle) is compared with that of the quantum kernel (train: green circle; test: blue circle). Note that the performance of our quantum kernel without introducing any hyperparameter is comparable to that of the classical kernel with the optimized bandwidth for the dimension $d$ smaller than $\sim 300$. Introducing the scaling parameter improved the performance of our quantum kernel for the dimension $d$ larger than $\sim 300$, which is indicated by the blue dotted line. (**b**) Test accuracies obtained by the quantum kernel for a range of features with varying the number of data samples ($N = 20, 100, 500, 1000, 4000$). The results were averaged over the three tasks: pullover vs. shirt (2 vs. 6), pullover vs. coat (2 vs. 4), and coat vs. shirt (4 vs 6) classification tasks. (**c**) Best test accuracies with respect to the number of data samples $N$, up to $N = 4000$ for the best classical (red triangle) and the quantum (blue circle) kernels. Each plot represents the best test accuracy for a given $N$. The results were averaged over the same three tasks as in (**b**). (**d**) The effect of quantum entanglement within a block (the block size $n = 2, 3, 6$). The coat vs. shirt (4 vs 6) classification task was used. The shaded regions indicate the standard deviation over 6 independent runs.

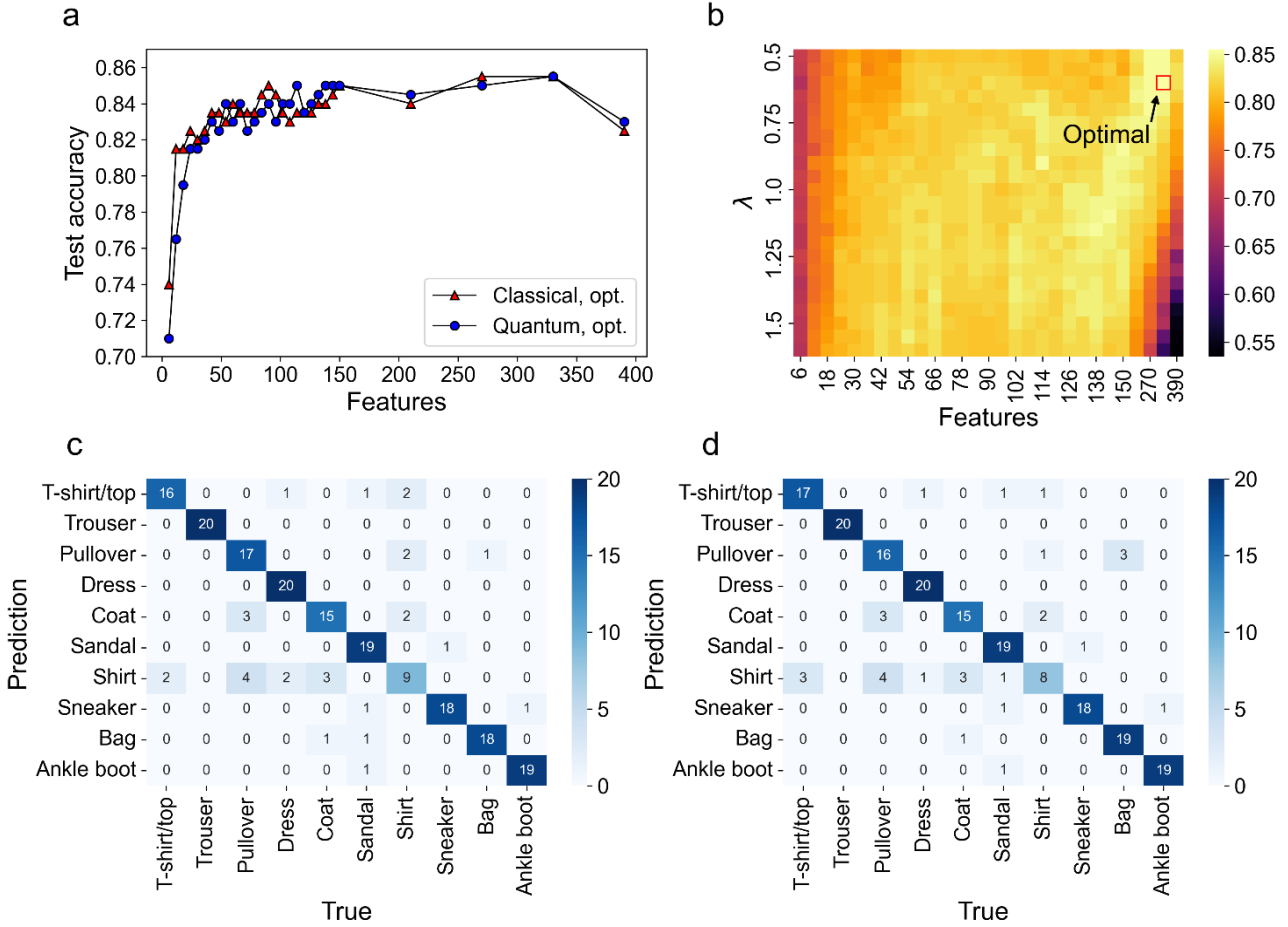

**Figure 4.** Multiclass classification on Fashion MNIST dataset. (**a**) Test accuracies for a range of PCA-reduced features from $d = 4$ to $d = 390$ (the block size $n = 2$) for the best classical (red triangle) and the quantum (blue circle) kernels. The number of data samples was 1000. For the classical kernel, we used Gaussian kernels with the optimized bandwidth for each dimension. For the quantum counterpart, the optimal scaling parameter $\lambda$ ($x^{(i)} \leftarrow \lambda x^{(i)}$) was used for each dimension. The quantum kernel with the optimal scaling parameter is competitive with the classical counterpart. (**b**) Grid search over the scaling parameter $\lambda$ for a range of features. The scaling parameter that gave the optimal test accuracy is indicated by the red open square: $\lambda^* = 0.6$ at $d = 330$. (**c**) Confusion matrix obtained from the best classical kernel (dimension $d = 330$). The performance metrics were as follows: accuracy, 0.855; precision, 0.853; recall, 0.855; F-measure, 0.851. (**d**) Confusion matrix obtained by the quantum kernel with the optimized scaling parameter (dimension $d = 330$). The performance metrics were as follows: accuracy, 0.855; precision, 0.850, recall, 0.855; F-measure, 0.848.

**Multiclass classification on Fashion-MNIST dataset.** We also show the numerical results for 10-class classification on Fashion-MNIST. We trained our multiclass QSVMs using a one-vs.-rest strategy. As was found in binary classification tasks, our quantum kernel was comparable to the best classical kernel (Fig. 4a). For multiclass classification using the quantum kernel, we found that it was important to introduce the scaling parameter. Hence, we performed a grid search for the scaling parameter $\lambda$ for a range of features ($4 < d < 340$) (Fig. 4b). The optimal value for $\lambda$ was 0.6 at $d = 330$. On the other hand, the optimal value for $\gamma$ of the Gaussian kernel was 2.5 at $d = 330$. The confusion matrices for QSVM and SVM were similar to each other (Fig. 4c–d). The performance metrics for the quantum (classical) kernel were the following: accuracy, 0.855 (0.855); precision, 0.850 (0.853); recall, 0.855 (0.855); F-measure, 0.848 (0.851). We note that, among 45 pairs generated by 10 categories of Fashion-MNIST, about half the pairs of classification tasks were relatively easy to distinguish; hence, the difference in the test accuracy between the classical and the quantum kernels tended to be decreased. The results suggest that our quantum kernel performed competitively with the best classical kernel in the multiclass classification task.

Discussion

In this study, we have implemented an application-specific quantum AI simulator using a heterogeneous CPU–FPGA computing, which was achieved by co-designing the FPGA architecture and our quantum kernel. To this end, we have introduced a BPS structure as a quantum feature map for QSVM, where a small number of qubits are entangled in each block. This is the first demonstration of the FPGA implementation of a gated-based quantum kernel. The co-design of the quantum kernel and its efficient FPGA implementation have enabled us to perform one of the largest numerical simulations of QSVM in terms of input features, up to 780-dimensional data. In the literature, one of the largest simulations of quantum kernels in terms of qubit count was performed by Huang *et al.* [11]. The number of qubits in their study is 30. For our particular study, increasing the number of entangled qubits is not a practical direction. Instead, our strategy is to divide input features into a number of blocks, and each block's quantum kernel can be efficiently computed in FPGA. By doing this, hundreds of features can be handled. Our approach is highly customized for our specific tasks at the hardware level; the focus of our simulator differs from that of a general-purpose quantum simulator, which is designed to be flexible and to perform various quantum algorithms. An application of our quantum kernel to dimensional features larger than ~1000 would be more challenging because off-diagonal kernel values could become much smaller. This limitation is related to our formalism of the quantum kernel, owing to the multiplication of many values that are less than one in Eq. (6). Nevertheless, the FPGA-based quantum kernel simulator has significantly accelerated our numerical simulations and allowed us to validate the applicability to QSVM with hundreds of input features. The quantum circuit presented in this work might have implications for co-designing quantum software and hardware and for developing application-specific quantum computers [42, 43].

      We have demonstrated that the FPGA-based quantum kernel simulator was 470 times faster than that obtained by the CPU implementation, without loss of accuracy. The numerical simulations show that our FPGA implementation is highly efficient in terms of the number of data samples (up to 4000), with a modest number of entangling qubits being used in the quantum feature map. We have applied our quantum kernel to image classification using Fashion-MNIST for a wide range of PCA-reduced features. The results

suggest that our quantum kernel is comparable to the best classical kernel, with similar generalization performance for binary and multiclass classification tasks. In binary classification, our hyperparameter-free quantum kernel was comparable to the Gaussian kernels; whereas, in multiclass classification, the scaling parameter played a significant role in improving the performance of our quantum kernel, in line with recent studies [15, 44].

Whether quantum kernels could perform better than classical kernels or have a practical advantage in real-world settings is still an open question. Our quantum kernel may be helpful for understanding the applicability of quantum kernels as well as their limitations. While our quantum kernel was applied to classification, the quantum kernel could be used for other kernel-based ML tasks, such as regression, spectral clustering, Gaussian process [17], and causal discovery [45]. With hundreds of input features being handled in our quantum kernel, other possible applications might include financial data, cheminformatics, and medical data. There is room for improvement in our quantum feature map. For instance, a recent approach based on the automatic design of quantum feature maps [46] may possibly improve our quantum feature map or reduce the number of quantum gates required. Nonetheless, our results might have implications for developing quantum-inspired algorithms and designing practical quantum kernels in the NISQ era.

## Methods

**FPGA implementation of the quantum kernel.** We describe an approach for efficient simulation of our quantum kernel, which is particularly designed for our FPGA architecture. The quantum kernel is given by the inner product of the quantum feature map, which in principle requires $\mathcal{O}(2^{3n})$ operations, owing to the multiplication of $2^n \times 2^n$ matrices to generate the quantum feature map. Such computational complexity becomes prohibitive for efficient FPGA implementation of quantum kernels, because FPGA architecture is memory-bound and the number of complex multipliers is limited. For that reason, efficient resource utilization of FPGA was crucial for calculating our quantum kernel. In this work, we employed a shallow quantum circuit so that we were able to calculate the quantum kernel with $\mathcal{O}(2^n)$ operations, as we will see below. This enabled efficient parallelization and the use of internal memory in FPGA.

We consider the following quantum state:

$$|\psi\rangle = (V_1 \otimes V_2 \otimes \cdots \otimes V_n) U_{2^n}^{\text{ent}} (U_1 \otimes U_2 \otimes \cdots \otimes U_n) |0^{\otimes n}\rangle. \tag{7}$$

where $U_1, U_2, \cdots, U_n$ and $V_1, V_2, \cdots, V_n$ are single-qubit gates and $U_{2^n}^{\text{ent}} := \prod_{q=1}^{n-1} \mathbf{CNOT}_{q,q+1}$ represents $n$-qubit entanglement operation. For the sake of our discussion, it is convenient to rewrite $|\psi\rangle$ as $\boldsymbol{f} = VU_{2^n}^{\text{ent}}U\boldsymbol{f}_0$ with $\boldsymbol{f}_0$ being a vector $[1,0,\cdots,0]^{\text{T}}$, where $U := U_1 \otimes \cdots \otimes U_n$ and $V := V_1 \otimes \cdots \otimes V_n$. First, we note that, in the calculation of $U\boldsymbol{f}_0$, only the first column of $U$ is needed; hence, $U\boldsymbol{f}_0$ can be obtained without the need for fully conducting tensor operations. By denoting the first column vector of each $2 \times 2$ unitary matrix $U_q$ as $\left[\chi_1^{(q)}, \chi_2^{(q)}\right]^{\text{T}}$ and the first column vector of $U$ as $\boldsymbol{u} = [u_1, u_2, \cdots, u_{2^n}]^{\text{T}} \in \mathbb{C}^{2^n}$, then we

have

$$Uf_0 = u = \begin{bmatrix} \chi_1^{(1)} \cdots \chi_1^{(n-2)} \cdot \chi_1^{(n-1)} \cdot \chi_1^{(n)} \\ \chi_1^{(1)} \cdots \chi_1^{(n-2)} \cdot \chi_1^{(n-1)} \cdot \chi_2^{(n)} \\ \chi_1^{(1)} \cdots \chi_1^{(n-2)} \cdot \chi_2^{(n-1)} \cdot \chi_1^{(n)} \\ \chi_1^{(1)} \cdots \chi_1^{(n-2)} \cdot \chi_2^{(n-1)} \cdot \chi_2^{(n)} \\ \vdots \\ \chi_2^{(1)} \cdots \chi_2^{(n-2)} \cdot \chi_1^{(n-1)} \cdot \chi_1^{(n)} \\ \chi_2^{(1)} \cdots \chi_2^{(n-2)} \cdot \chi_1^{(n-1)} \cdot \chi_2^{(n)} \\ \chi_2^{(1)} \cdots \chi_2^{(n-2)} \cdot \chi_2^{(n-1)} \cdot \chi_1^{(n)} \\ \chi_2^{(1)} \cdots \chi_2^{(n-2)} \cdot \chi_2^{(n-1)} \cdot \chi_2^{(n)} \end{bmatrix}.$$

(8)

This calculation can be performed by $4 \cdot (2^{n-1} - 1)$ operations using complex multipliers in FPGA (more details are given in Supplementary Figure 5). The feature map can thus be rewritten as $f = V U_{2^n}^{\text{ent}} u$. Next, we note that $V$ is a diagonal matrix in our quantum circuit and that $U_{2^n}^{\text{ent}}$ is a sparse matrix, in which each row vector contains only one non-zero entry. By denoting the diagonal elements $\{V_{kk}\}$ as $v = [v_1, v_2, \cdots, v_{2^n}]^T \in \mathbb{C}^{2^n}$, we can calculate $f$ as

$$f_k = v_k u_{\xi_k}.$$

(9)

Here $\xi_k$ is the index of the non-zero element in the $i$th row of $U_{2^n}^{\text{ent}}$ (e.g., for $n = 2$, then $\xi_1 = 1$, $\xi_2 = 2$, $\xi_3 = 4$, and $\xi_4 = 3$). In general, $U_{2^n}^{\text{ent}}$ can be calculated recursively by

$$U_{2^{n+1}}^{\text{ent}} = \begin{bmatrix} U_{2^n}^{\text{ent}} & O_{2^n} \\ O_{2^n} & Y_{2^n} \end{bmatrix}; \quad Y_{2^{n+1}} = \begin{bmatrix} O_{2^n} & U_{2^n}^{\text{ent}} \\ Y_{2^n} & O_{2^n} \end{bmatrix} \quad (n \geq 1).$$

(10)

where $U_2^{\text{ent}}$ and $Y_2$ denote the $2 \times 2$ identity matrix and the Pauli X matrix, respectively, and $O_{2^n}$ denotes the $2^n \times 2^n$ zero matrix. The proof of the recurrence relation is given in Supplementary Note 1. The indices $\{\xi_k\}$ in Eq. (9) can be determined once $U_{2^n}^{\text{ent}}$ is obtained. Finally, the inner product $\langle \psi^i | \psi^j \rangle$ can be calculated by $\sum_k f_k^*(s^{(i)}) f_k(s^{(j)})$.

Details of computational resources. Our quantum AI simulator based on a hybrid CPU–FPGA system is implemented on the Amazon Web Services (AWS) Elastic Computing Cloud (EC2) platform, in which AWS EC2 F1 instances of AMD Xilinx FPGA hardware are accessible. In particular, we used the f1.2xlarge instance size, which has 1 FPGA, 8 vCPUs, and 122 GB of memory. More specifically, we used AMD Xilinx Virtex™ UltraScale+™ VU19P FPGA and Intel Xeon™ E5-2686 v4 with a base clock speed of 2.3 GHz as vCPU. The details of our FPGA architecture and block diagrams are provided in Supplementary Note 2.

Machine learning. Here we provide the details of our ML models. Preprocessing was applied to the original data to make them suitable for quantum angle encoding: PCA was used to reduce the dimension of the $28 \times 28$ original image data to $d$-dimensional input vectors $x^{(i)} \in \mathbb{R}^d$ (where $d$ was varied from 4 to 780), which were then transformed such that $x^{(i)} \in [-1,1]$. In the training of support vector classifiers, hinge loss was used for the loss function. Throughout the paper, the regularization parameter $C$ for soft margin SVM [47] was set to 1.0 for both classical and quantum ML models. For the multiclass classification task shown in Fig. 4, a one-vs.-rest strategy was employed.

To compare the performance of our quantum kernel with the classical counterpart, we used the Gaussian kernel, which is given by $\exp\left(-\gamma \|x^{(i)} - x^{(j)}\|^2\right)$, with $\gamma$ being a hyperparameter. To obtain the optimal test accuracy, we performed a grid search over the bandwidth

$$\gamma \in \{0.001, 0.1, 0.25, 0.5, 0.75, 1, 1.25, 2.5, 3.75, 5, 6.25, 7.5, 8.75, 10, 50, 100, 1000\}. \tag{11}$$

It is also possible to introduce a hyperparameter in our quantum feature map $|\Psi^{\text{BPS}}(x)\rangle$. In this work, we consider that the input vector $x$ can be scaled by $\lambda$ (i.e., $x^{(i)} \leftarrow \lambda x^{(i)}$), which is similar to an approach by recent work [15]. Thus, we performed a grid search over the scaling parameter

$$\lambda \in \{0.001, 0.1, 0.25, 0.5, 0.75, 1, 1.25, 2.5, 3.75, 5, 6.25, 7.5, 8.75, 10, 50, 100, 1000\}. \tag{12}$$

The effect of the scaling parameter $\lambda$ was somewhat different from that of $\gamma$. In particular, we found that, for binary classification, the case of $\lambda = 1$ typically gave the near-optimal performance (see also Supplementary Note 3), implying that our quantum kernel gave a reasonable performance without introducing any hyperparameter. Nonetheless, to further optimize the value for $\lambda$, we narrowed the range for $\lambda$ and performed another grid search over the scaling parameter

$$\lambda \in \{0.5, 0.55, 0.6, 0.65, 0.7, 0.75, 0.8, 0.85, 0.9, 0.95, 1, 1.05,$$
$$1.1, 1.15, 1.2, 1.25, 1.3, 1.35, 1.4, 1.45, 1.5, 1.55, 1.6\}. \tag{13}$$

We found that the test accuracy was slightly improved from 0.870 to 0.875 in binary classification (see also Supplementary Note 4) and that the use of the scaling parameter $\lambda$ played an important role in multiclass classification.

Data availability

All the datasets used in the present study are publicly available at https://github.com/zalandoresearch/fashion-mnist. We cited the reference of the data in the manuscript.

References
[1] Nielsen, M. A. & Chuang, I. L. *Quantum Computing and Quantum Information, 10th Anniversary Ed.*

(Cambridge University Press, 2010).

[2] Woerner, S. & Egger, D. J. Quantum risk analysis. *npj Quantum Inf.* **5**, 15. https://doi.org/10.1038/s41534-019-0130-6 (2019).

[3] Cao, Y. *et al.* Quantum chemistry in the age of quantum computing. *Chem. Rev.* **119**, 10856–10915 (2019).

[4] Rebentrost, P., Mohseni, M. & Lloyd, S. Quantum support vector machine for big data classification. *Phys. Rev. Lett.* **113**, 130503 (2014).

[5] Liu, Y., Arunachalam, S. & Temme, K. A rigorous and robust quantum speed-up in supervised machine learning. *Nat. Phys.* **17**, 1013–1017 (2021).

[6] Biamonte, J. *et al.* Quantum machine learning. *Nature* **549**, 195–202 (2017).

[7] Mitarai, K., Negoro, M., Kitagawa, M. & Fujii, K. Quantum circuit learning. *Phys. Rev. A* **98**, 032309 (2018).

[8] Schuld, M. & Killoran, N. Quantum machine learning in feature Hilbert spaces. *Phys. Rev. Lett.* **122**, 040504 (2019).

[9] Havlíček, V. *et al.* Supervised learning with quantum-enhanced feature spaces. *Nature* **567**, 209–212 (2019).

[10] Benedetti, M., Lloyd, E., Sack, S. & Fiorentini, M. Parameterized quantum circuits as machine learning models. *Quantum Sci. Technol.* **4**, 043001 (2019).

[11] Huang, H.-Y. *et al.* Power of data in quantum machine learning. *Nat. Commun.* **12**, 2631. https://doi.org/10.1038/s41467-021-22539-9 (2021).

[12] Peters, E. *et al.* Machine learning of high dimensional data on a noisy quantum processor. *npj Quantum Inf.* **7**, 161. https://doi.org/10.1038/s41534-021-00498-9 (2021).

[13] Hubregtsen, T. *et al.* Training quantum embedding kernels on near-term quantum computers. Preprint at https://arxiv.org/abs/2105.02276 (2021).

[14] Jerbi, S. *et al.* Quantum machine learning beyond kernel methods. Preprint at https://arxiv.org/abs/2110.13162 (2021).

[15] Shaydulin, R. & Wild, S. M. Importance of kernel bandwidth in quantum machine learning. Preprint at https://arxiv.org/abs/2111.05451 (2021).

[16] Kusumoto, T., Mitarai, K., Fujii, K., Kitagawa, M. & Negoro, M. Experimental quantum kernel trick with nuclear spins in a solid. *npj Quantum Inf.* **7**, 94. https://doi.org/10.1038/s41534-021-00423-0 (2021).

[17] Moradi, S. *et al.* Error mitigation for quantum kernel based machine learning methods on IonQ and IBM quantum computers. Preprint at https://arxiv.org/abs/2206.01573 (2022).

[18] Postler, L. *et al.* Demonstration of fault-tolerant universal quantum gate operations. *Nature* **605**, 675–680 (2022).

[19] Preskill, J. Quantum computing in the NISQ era and beyond. *Quantum* **2**, 79. https://doi.org/10.22331/q-2018-08-06-79 (2018).

[20] Bharti, K. *et al.* Noisy intermediate-scale quantum algorithms. *Rev. Mod. Phys.* **94**, 015004 (2022).

[21] Aleksandrowicz, G. *et al.* Qiskit: An open-source framework for quantum computing. https://doi.org/10.5281/ZENODO.2562111 (2019).

[22] Guo, C. *et al.* General-purpose quantum circuit simulator with projected entangled-pair states and the quantum supremacy frontier. *Phys. Rev. Lett.* **123**, 190501 (2019).

[23] Wang, Z. *et al.* A quantum circuit simulator and its applications on Sunway TaihuLight supercomputer. *Sci. Rep.* **11**, 355. https://doi.org/10.1038/s41598-020-79777-y (2021).

[24] Suzuki, Y. *et al.* Qulacs: a fast and versatile quantum circuit simulator for research purpose. *Quantum*, **5**, 559. https://doi.org/10.22331/q-2021-10-06-559 (2021).

[25] Efthymiou, S. *et al.* Qibo: a framework for quantum simulation with hardware acceleration. *Quantum Sci. Technol.* **7**, 015018 (2022).

[26] Vincent, T. *et al.* Jet: fast quantum circuit simulations with parallel task-based tensor-network contraction. *Quantum* **6**, 709. https://doi.org/10.22331/q-2022-05-09-709 (2022).

[27] Nguyen, T. *et al.* Tensor network quantum virtual machine for simulating quantum circuits at exascale. Preprint at https://arxiv.org/abs/2104.10523 (2021).

[28] Khalid, A. U., Zilic, Z. & Radecka, K. FPGA emulation of quantum circuits. In *IEEE International Conference on Computer Design: VLSI in Computers and Processors (ICCD)* (2004).

[29] Lee, Y. H., Khalil-Hani, M. & Marsono, M. N. An FPGA-based quantum computing emulation framework based on serial-parallel architecture. *Int. J. Reconfigurable Comput.* **2016**, 5718124 (2016).

[30] Mahmud, N. & El-Araby, E. A scalable high-precision and high-throughput architecture for emulation of quantum algorithms. In *2018 31st IEEE International System-on-Chip Conference (SOCC)* (2018).

[31] Mahmud, N., El-Araby, E. & Caliga, D. Scaling reconfigurable emulation of quantum algorithms at high precision and high throughput. *Quantum Eng.* **1**, e19. https://doi.org/10.1002/que2.19 (2019).

[32] Pilch, J. & Długopolski, J. An FPGA-based real quantum computer emulator. *J. Comput. Electron.* **18**, 329–342 (2019).

[33] Cortes, C. & Vapnik, V. Support-vector networks. *Mach. Learn.* **20**, 273–297 (1995).

[34] Schölkopf, B. & Smola, A. J. *Learning with Kernels: Support Vector Machines, Regularization, Optimization, and Beyond.* (MIT Press, 2002).

[35] Bremner, M. J., Jozsa, R. & Shepherd, D. J. Classical simulation of commuting quantum computations implies collapse of the polynomial hierarchy. *Proc. R. Soc. A* **467**, 459 (2011).

[36] Martyn, J., Vidal, G., Roberts, C. & Leichenauer, S. Entanglement and tensor networks for supervised image classification. Preprint at https://arxiv.org/abs/2007.06082 (2020).

[37] Suzuki, T. & Katouda, M. Predicting toxicity by quantum machine learning. *J. Phys. Commun.* **4**, 125012. https://doi.org/10.1088/2399-6528/abd3d8 (2020).

[38] Subasi, A. & Gursoy, M. I. EEG signal classification using PCA, ICA, LDA and support vector machines. *Expert Syst. Appl.* **37**, 8659–8666 (2010).

[39] Xiao, H., Rasul, K. & Vollgraf, R. Fashion-MNIST: a novel image dataset for benchmarking machine learning algorithms. Preprint at https://arxiv.org/abs/1708.07747 (2017).

[40] Volder, J. E. The CORDIC trigonometric computing technique. *IRE Trans. Electron. Comput.* **3**, 330–334 (1959).

[41] Harris, C. R. *et al.* Array programming with NumPy. *Nature* **585**, 357–362 (2020).

[42] Li, G. *et al.* On the co-design of quantum software and hardware. In *Proceedings of the Eight Annual ACM International Conference on Nanoscale Computing and Communication* (2021).

[43] Tomesh, T. & Martonosi, M. Quantum codesign. *IEEE Micro*. **41**, 33–40 (2021).

[44] Canatar, A., Peters, E., Pehlevan, C., Wild, S. M. & Shaydulin, R. Bandwidth enables generalization in quantum kernel models. Preprint at https://arxiv.org/abs/2206.06686 (2022).

[45] Kawaguchi, H. Application of quantum computing to a linear non-Gaussian acyclic model for novel medical knowledge discovery. Preprint at https://arxiv.org/abs/2110.04485 (2021).

[46] Altares-López, S., Ribeiro, A. & García-Ripoll, J. J. Automatic design of quantum feature maps. *Quantum. Sci. Technol.* **6**, 045015 (2021).

[47] Chang, C. C. & Lin, C. J. LIBSVM: a library for support vector machines. *ACM Trans. Intell. Syst. Technol.* **2**, 1–27 (2011).

Acknowledgment

We thank Hideki Asoh (National Institute of Advanced Industrial Science and Technology) for useful discussions.

Author contributions

T.S. and T.O. conceived the concept of co-designing the quantum kernel and the FPGA implementation. T.S. conceived the idea of the quantum kernel for image classification in this work. T.M. developed the computer code for the quantum kernel simulator. T.I. executed the FPGA implementation and performed the numerical simulations. T.S., T.I., and T.M. analyzed the results of the quantum support vector machines. T.S. wrote the manuscript. All the authors commented on the manuscript.

**Correspondence** and requests for materials should be addressed to T.S.

Supplementary Information for

"Quantum AI simulator using a hybrid CPU–FPGA approach"

Teppei Suzuki[1,*], Tsubasa Miyazaki[1], Toshiki Inaritai[1], and Takahiro Otsuka[1]

[1] *Research and Development Center, SCSK Corporation, Toyosu Front, 3-2-20 Toyosu, Koto-ku, Tokyo 135-8110, Japan*

Supplementary Note 1. Recurrence relation for an $n$-qubit entanglement operation matrix

We discuss how we can calculate a unitary matrix $U_{2^n} = \prod_{q=1}^{n-1} \mathbf{CNOT}_{q,q+1}$, without actually conducting tensor product operations. The matrix $U_{2^n}$ is a sparse matrix that can be recursively obtained using Proposition 1.

**Proposition 1** (Recurrence relation). *Let $n \in \mathbb{N}$. Let $\{U_{2^n}\}$ and $\{Y_{2^n}\}$ be sequences of square matrices such that*

$$U_{2^{n+1}} := (I_{2^{n-1}} \otimes \mathbf{CNOT})(U_{2^n} \otimes \mathbf{ID}),$$
$$Y_{2^{n+1}} := (I_{2^{n-1}} \otimes \mathbf{CNOT})(Y_{2^n} \otimes \mathbf{ID}),$$

(A1)

*where $\mathbf{CNOT}$ and $\mathbf{ID}$ are the matrices representing the controlled NOT gate and the identity gate, respectively and $I_{2^n}$ denotes the $2^n \times 2^n$ identity matrix, with $I_{2^0} := 1$; and let $U_2$ and $Y_2$ be defined by the $2 \times 2$ identity matrix and the Pauli X matrix, respectively:*

$$U_2 := I_2 = \begin{bmatrix} 1 & 0 \\ 0 & 1 \end{bmatrix}; \quad Y_2 := X = \begin{bmatrix} 0 & 1 \\ 1 & 0 \end{bmatrix}.$$

(A2)

*Then $U_{2^{n+1}}$ and $Y_{2^{n+1}}$ can be calculated by recursion*

$$U_{2^{n+1}} = \begin{bmatrix} U_{2^n} & O_{2^n} \\ O_{2^n} & Y_{2^n} \end{bmatrix} \quad (n \geq 1),$$

$$Y_{2^{n+1}} = \begin{bmatrix} O_{2^n} & U_{2^n} \\ Y_{2^n} & O_{2^n} \end{bmatrix} \quad (n \geq 1),$$

(A3)

*where $O_{2^n}$ denotes the $2^n \times 2^n$ zero matrix.*

*Proof.* We prove the statement by induction on $n$.
Step I. Base case ($n = 1$):

$$U_4 = (1 \otimes \mathbf{CNOT})(U_2 \otimes \mathbf{ID}) = (1 \otimes \mathbf{CNOT})(I_2 \otimes \mathbf{ID}) = \begin{bmatrix} I_2 & O \\ O & X \end{bmatrix}\begin{bmatrix} I_2 & O \\ O & I_2 \end{bmatrix} = \begin{bmatrix} U_2 & O_2 \\ O_2 & Y_2 \end{bmatrix},$$

(A4)

and

$$Y_4 = (1 \otimes \mathbf{CNOT})(Y_2 \otimes \mathbf{ID}) = (1 \otimes \mathbf{CNOT})(X \otimes \mathbf{ID}) = \begin{bmatrix} I_2 & O \\ O & X \end{bmatrix}\begin{bmatrix} O & I_2 \\ I_2 & O \end{bmatrix} = \begin{bmatrix} O & I_2 \\ X & O \end{bmatrix} = \begin{bmatrix} O_2 & U_2 \\ Y_2 & O_2 \end{bmatrix}.$$

(A5)

Hence, the statement is true for $n = 1$. Note that $U_4$ is the **CNOT** gate itself.

Step II. Induction step: we assume that the statement holds for some natural number $k$. For $n = k+1$, we have

$$\begin{aligned}
U_{2^{k+2}} &= (I_{2^k} \otimes \mathbf{CNOT})(U_{2^{k+1}} \otimes \mathbf{ID}) \\
&= \begin{bmatrix} I_{2^{k-1}} \otimes \mathbf{CNOT} & O_{2^{k+1}} \\ O_{2^{k+1}} & I_{2^{k-1}} \otimes \mathbf{CNOT} \end{bmatrix}\begin{bmatrix} U_{2^k} \otimes \mathbf{ID} & O_{2^{k+1}} \\ O_{2^{k+1}} & Y_{2^k} \otimes \mathbf{ID} \end{bmatrix} \\
&= \begin{bmatrix} (I_{2^{k-1}} \otimes \mathbf{CNOT})(U_{2^k} \otimes \mathbf{ID}) & O_{2^{k+1}} \\ O_{2^{k+1}} & (I_{2^{k-1}} \otimes \mathbf{CNOT})(Y_{2^k} \otimes \mathbf{ID}) \end{bmatrix} = \begin{bmatrix} U_{2^{k+1}} & O_{2^{k+1}} \\ O_{2^{k+1}} & Y_{2^{n+1}} \end{bmatrix},
\end{aligned}$$

(A6)

and

$$\begin{aligned}
Y_{2^{k+2}} &= (I_{2^k} \otimes \mathbf{CNOT})(Y_{2^{k+1}} \otimes \mathbf{ID}) \\
&= \begin{bmatrix} I_{2^{k-1}} \otimes \mathbf{CNOT} & O_{2^{k+1}} \\ O_{2^{k+1}} & I_{2^{k-1}} \otimes \mathbf{CNOT} \end{bmatrix}\begin{bmatrix} O_{2^{k+1}} & U_{2^k} \otimes \mathbf{ID} \\ Y_{2^k} \otimes \mathbf{ID} & O_{2^{k+1}} \end{bmatrix} \\
&= \begin{bmatrix} O_{2^{k+1}} & (I_{2^{k-1}} \otimes \mathbf{CNOT})(U_{2^k} \otimes \mathbf{ID}) \\ (I_{2^{k-1}} \otimes \mathbf{CNOT})(Y_{2^k} \otimes \mathbf{ID}) & O_{2^{k+1}} \end{bmatrix} = \begin{bmatrix} O_{2^{k+1}} & U_{2^{k+1}} \\ Y_{2^{k+1}} & O_{2^{k+1}} \end{bmatrix}.
\end{aligned}$$

(A7)

We can see that the statement holds true for $n = k+1$. Hence, the statement holds for all natural numbers $n \geq 1$. $\square$

Note that matrices $U_{2^n}$ and $Y_{2^n}$ are sparse; and the sparsity pattern visualization for $U_{2^n}$ and $Y_{2^n}$ is shown Supplementary Figure 1. In the calculation of our quantum feature map, we only need the index for non-zero entry in each row vector of $U_{2^n}$.

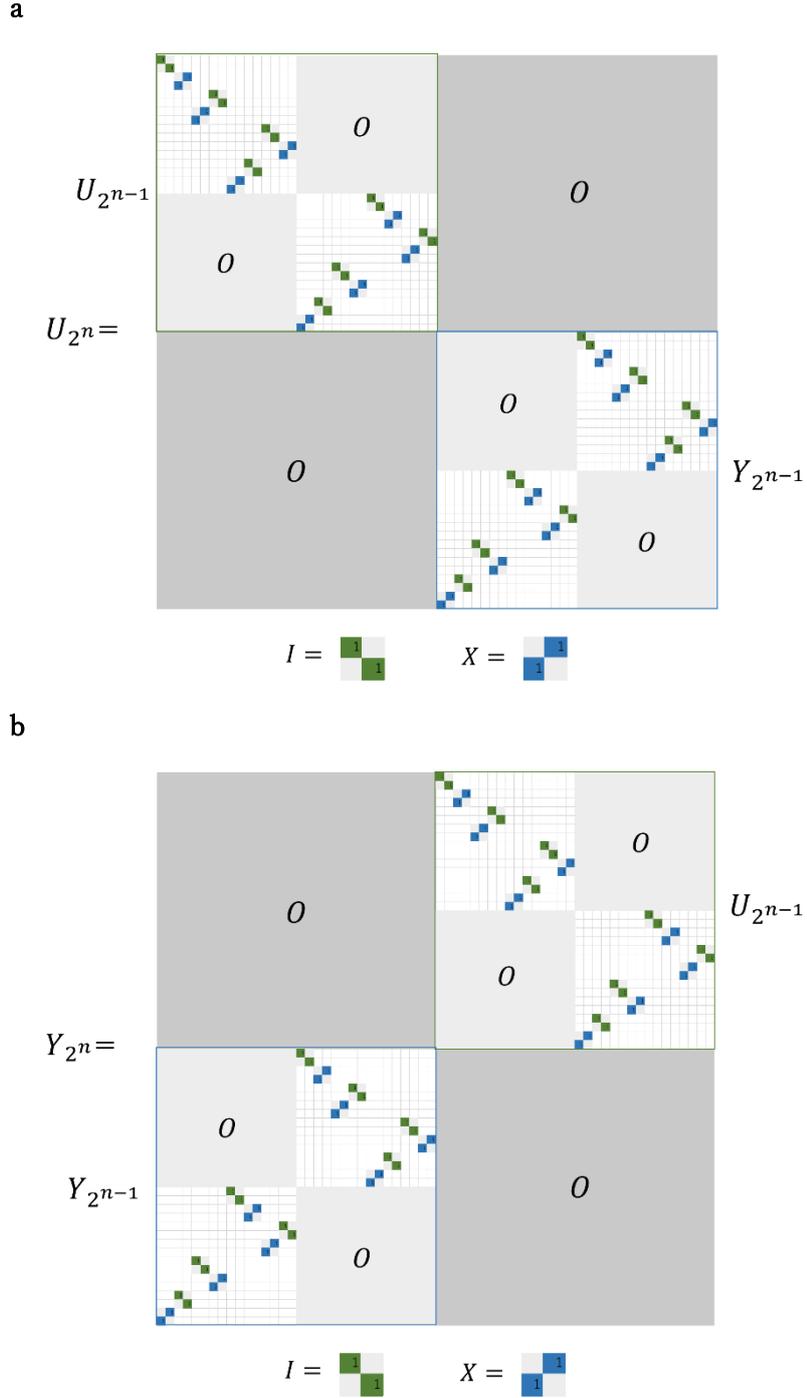

Supplementary Figure 1: Sparsity pattern visualization for $U_{2^n}$ and $Y_{2^n}$. Shown are matrices $U_{2^n}$ (a) and $Y_{2^n}$ (b) in the case of $n = 6$ (i.e., $U_{64}$ and $Y_{64}$). Values of one that are originally belonging to matrices $I$ and $X$ are represented by green and blue, respectively, while other entries are zero. The matrix $U_{2^n}$ represents an $n$-qubit entanglement operation $\prod_{q=1}^{n-1} \text{CNOT}_{q,q+1}$. The matrix $U_{2^n}$ has a property $\text{Tr}(U_{2^n}) = 2$ for all $n$.

## Supplementary Note 2. FPGA architecture and block diagrams for computing the quantum kernel

In this section, we describe the details of our FPGA architecture and block diagrams for computing the quantum kernel. The overview of the quantum kernel implementation is shown in Supplementary Figure 2. Our quantum AI simulator based on a hybrid CPU–FPGA approach is implemented on Amazon Web Services cloud platform, in which Amazon EC2 F1 instances of Xilinx FPGA hardware are accessible. First, PCA-reduced features are sent from CPU (the host) to FPGA via PCIe. Second, the sine and cosine of the input angles are computed using the CORDIC algorithm [S1] (Supplementary Figure 3). Third, the unitary matrices $U$ and $V$ (which are defined by Eq. (7) in the text) are computed (Supplementary Figures 4 and 5). Fourth, the quantum feature map is calculated using the unitary matrices $U$ and $V$, as well as an efficient implementation of $n$-qubit quantum entanglement (Supplementary Figure 6). Fifth, the square of the norm of the inner product is obtained (Supplementary Figure 7). Finally, the data is sent back to the host (CPU). In Supplementary Table 1, we give the details hardware utilizations for the quantum kernel implementation.

- Supplementary Figure 2: Scheme for the quantum kernel implementation.
- Supplementary Figure 3: Schematic block diagram for the module that computes sine and cosine functions of the input angles.
- Supplementary Figure 4: Block diagram for the module that computes the unitary matrices $U_q$ and $V_q$ using the sine and the cosine values.
- Supplementary Figure 5: Schematic block diagram for the module that computes the tensor product.
- Supplementary Figure 6: Schematic block diagram for the module that computes the quantum feature map.
- Supplementary Figure 7: Block diagram for the module that computes the square of the norm of the inner product.
- Supplementary Table 1: Hardware utilizations for the quantum kernel implementation.

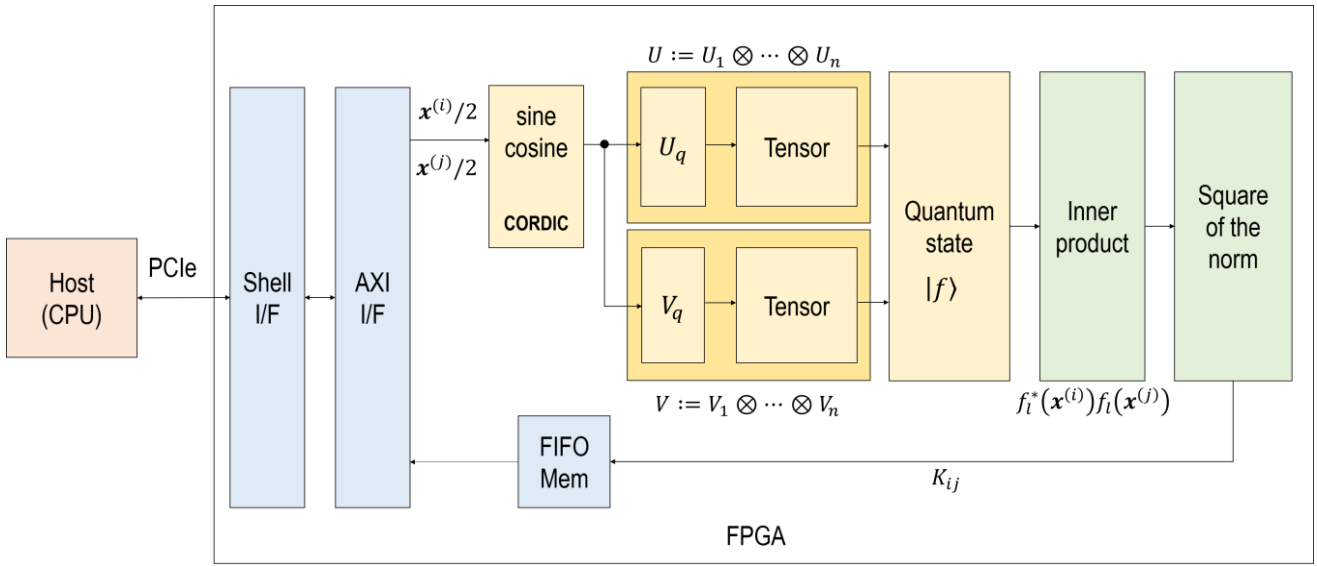

Supplementary Figure 2: Scheme for the quantum kernel implementation. Our quantum AI simulator based on a hybrid CPU–FPGA system is implemented on Amazon Web Services cloud platform, in which Amazon EC2 F1 instances of Xilinx FPGA hardware are accessible. PCA-reduced features are sent from CPU (the host application) to FPGA via PCIe. The quantum feature map can be obtained in the following steps (denoted by the yellow boxes): First, the sine and cosine of the input angles are computed using the CORDIC algorithm. Second, the unitary matrices $U$ and $V$ (which are defined by Eq. (7) in the text) are computed. Third, the quantum feature map is calculated using the unitary matrices $U$ and $V$, as well as an efficient implementation of $n$-qubit quantum entanglement. The square of the norm of the inner product is computed (denoted by the green boxes), generating the quantum kernel entry $K_{ij}$. Finally, the data is sent back to the host.

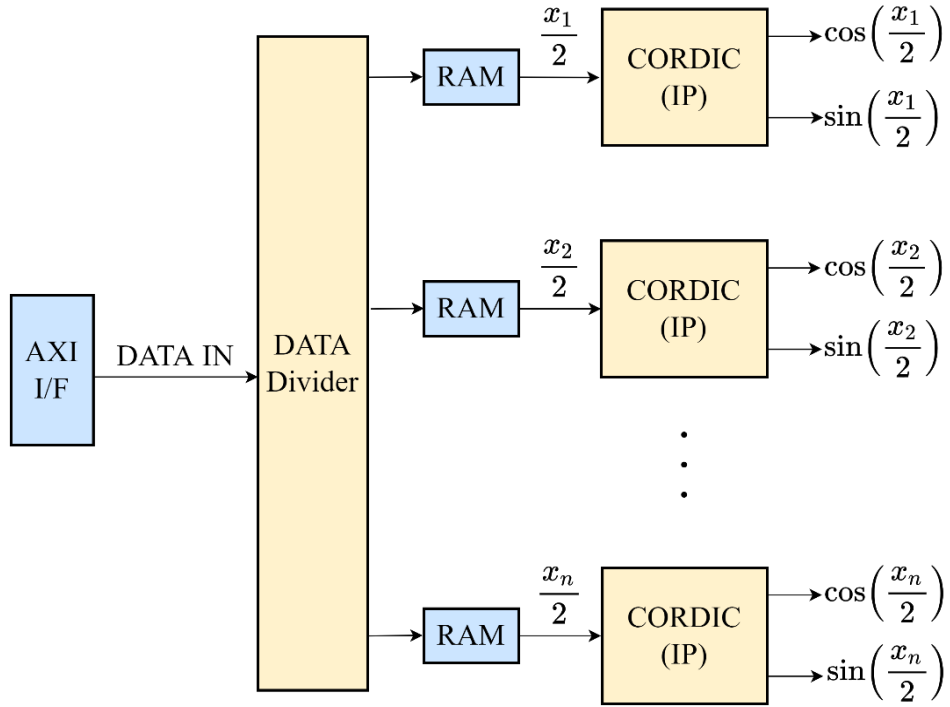

Supplementary Figure 3: Schematic block diagram for the module that computes sine and cosine functions of the input angles. Data input (PCA-reduced features) in CPU are sent and recognized as a streaming data by AXI, which is an interface between the host application (CPU) and FPGA. The data are divided into units consisting of $\frac{x_1}{2}, \frac{x_2}{2}, \cdots, \frac{x_n}{2}$ (via the *data divider*); and each value is stored in RAM. Then the stored data are read out and the sine and cosine of the input angles are computed using the CORDIC algorithm.

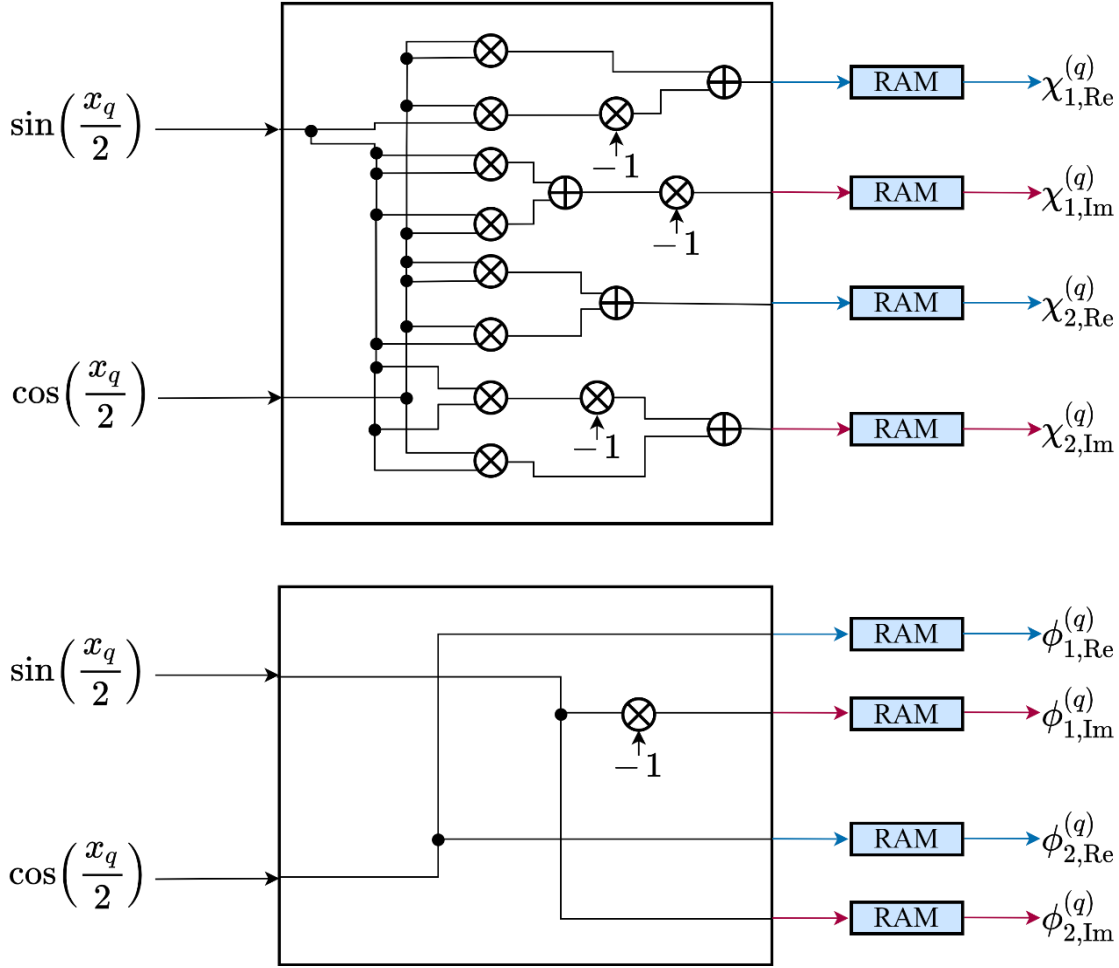

Supplementary Figure 4: Block diagram for the module that computes the unitary matrices $U_q$ and $V_q$ using the sine and the cosine values. (Top) Complex-valued entries $\chi_1^{(q)}$ and $\chi_2^{(q)}$ are generated using the sine and cosine of the input data in order to obtain the first column vector of each $2 \times 2$ unitary matrix $U_q = R_y(x_q) R_z(x_q) H$; the prefactor associated to the Hadamard gate will be multiplied later in the calculation. (Bottom) Complex-valued entries $\phi_1^{(q)}$ and $\phi_2^{(q)}$ are generated in order to obtain the diagonal elements of each $2 \times 2$ unitary matrix $V_q = R_z(x_q)$. Re and Im in the figure denote the real and imaginary parts of a complex number, respectively.

**a**

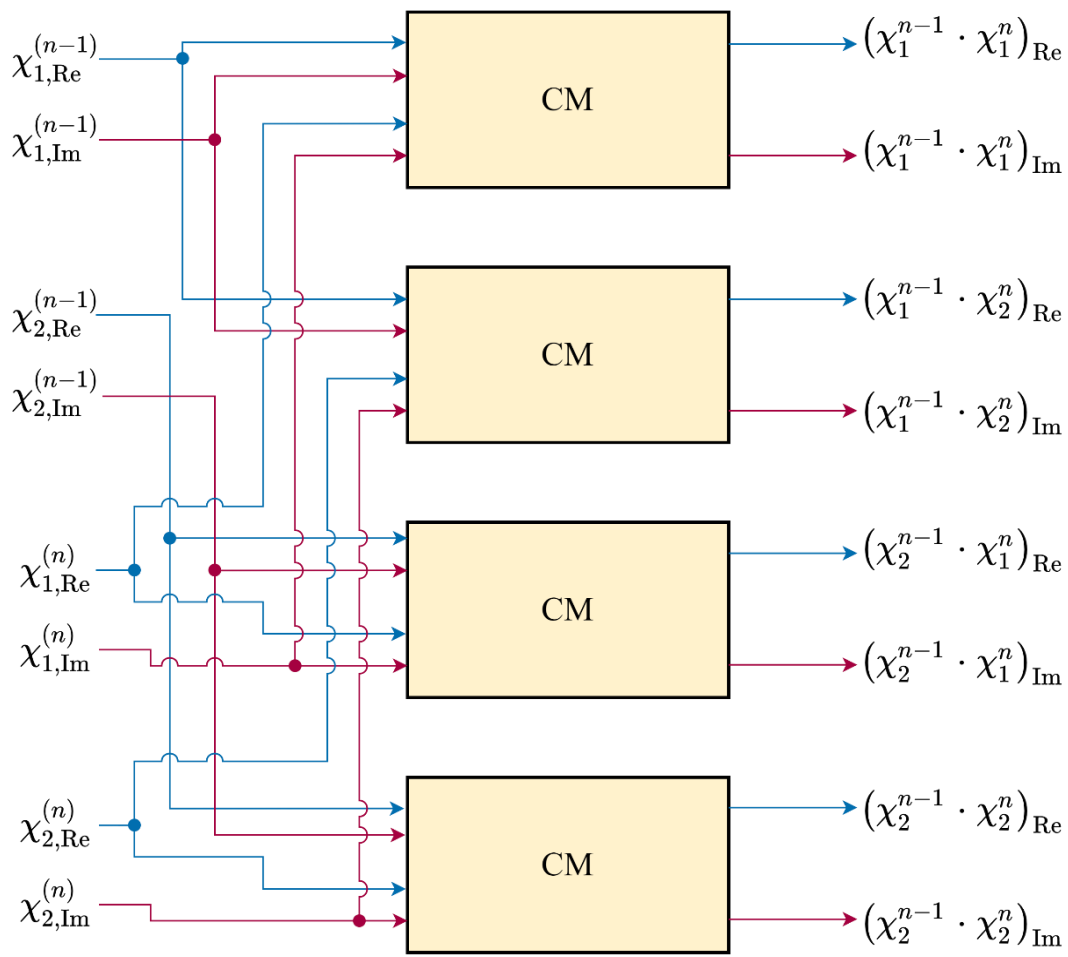

**b**

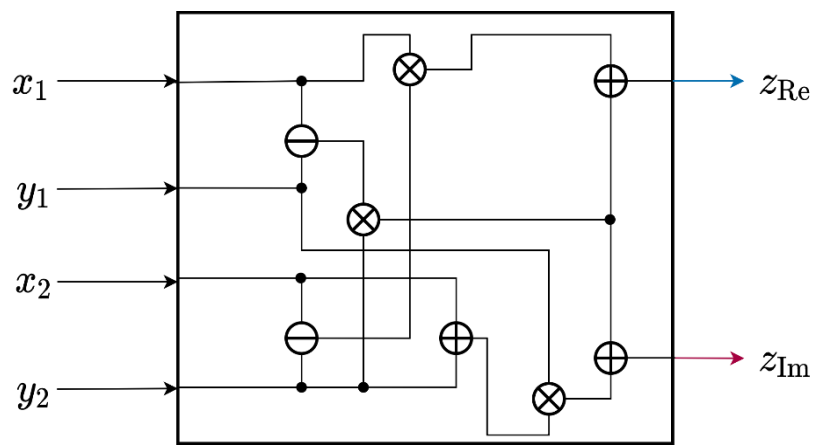

Supplementary Figure 5 (*continued*).

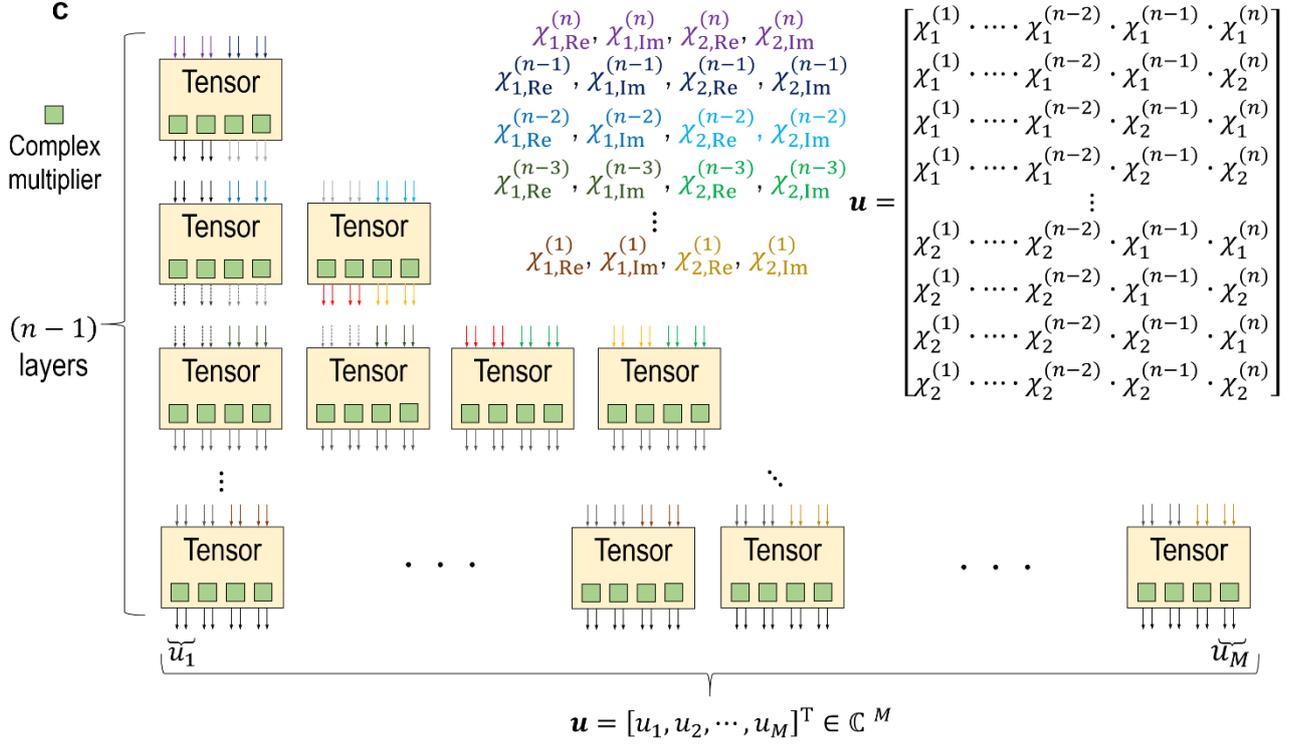

Supplementary Figure 5: Schematic block diagram for the module that computes the tensor product. Re and Im denote the real and imaginary parts of a complex number, respectively. (a) Schematic block diagram for a tensor-product subunit. Shown is an example for the input $\left(\chi_1^{(n-1)}, \chi_2^{(n-1)}\right)$ and $\left(\chi_1^{(n)}, \chi_2^{(n)}\right)$, which is the first layer in panel c. CM denotes complex multiplier, which is given in panel b. (b) Block diagram for complex multiplier: $z = (x_1 + iy_1) \cdot (x_2 + iy_2)$. (c) Schematic representation of the tensor product operation in FPGA. The tensor-product subunits are denoted by the yellow boxes, each of which contains four complex multipliers (see also part a). In our quantum feature map, only the first column of the matrix $U = U_1 \otimes \cdots \otimes U_n$ is needed; thus, it can be computed by Eq. (8), which is also shown in the right corner of panel c. From the schematic diagram depicted by panel c, we can see that the number of complex multipliers for this module is $4 \cdot (2^{n-1} - 1)$. During the calculation, a prefactor of $1/2$, which is associated to the Hadamard gate, is multiplied every two steps (if the total number of the steps is odd, then a prefactor of $1/2 = (1/\sqrt{2}) \cdot (1/\sqrt{2})$ is multiplied at the very end of the calculation of the square of the norm of the inner product (see also Supplementary Figure 7)). In a similar manner, we can also obtain $V = V_1 \otimes \cdots \otimes V_n$, where $V_q$ is given by the rotation operator gate $R_z(x_q)$; in this case, only the operations involving the diagonal elements of $V$ are needed and there is no need for the prefactor adjustment.

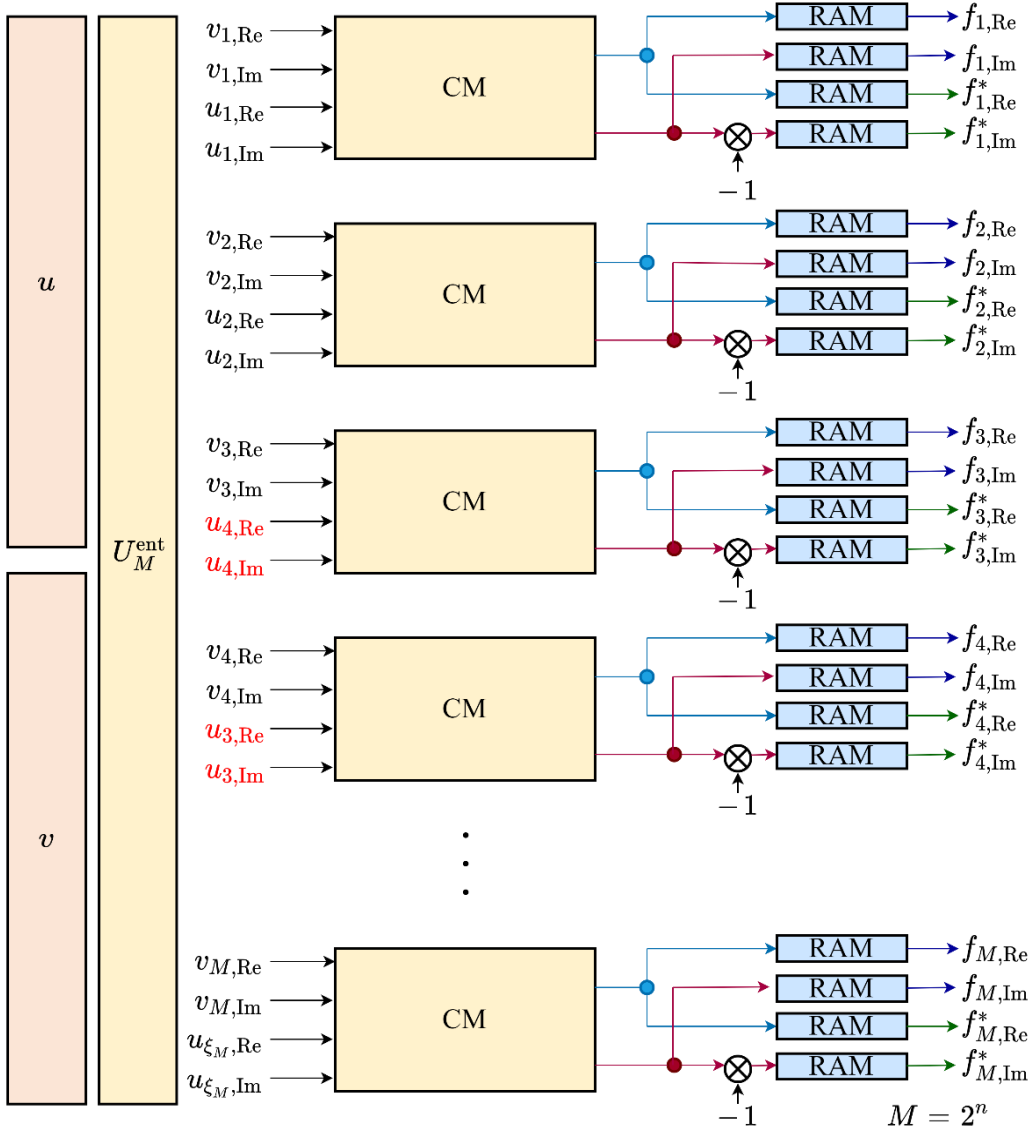

Supplementary Figure 6: Schematic block diagram for the module that computes the quantum feature map. On the basis of Eq. (9), the quantum state $\boldsymbol{f} \in \mathbb{C}^{2^n}$ can be computed using the unitary matrices $U$ and $V$ (which are obtained by the modules described in Supplementary Figures 4 and 5). The vector $\boldsymbol{u}$ is the first column of the unitary matrix $U$; and the vector $\boldsymbol{v}$ is the diagonal element of the diagonal matrix $V$. In our quantum feature map defined by Eq. (7), the role of $n$-qubit entanglement $\prod_{q=1}^{n-1} \mathbf{CNOT}_{q,q+1}$ can be viewed as 'rearranging' the elements of the vector $\boldsymbol{u}$ in accordance with the matrix $U_{2^n}$ defined in Supplementary Figure 1a. For instance, $f_3 = v_3 u_{\xi_3} = v_3 u_4$ and $f_4 = v_4 u_{\xi_4} = v_4 u_3$, which are indicated by red in the figure. Such technique leads to an efficient implementation of the quantum state including quantum entanglement. The complex conjugate of the state vector $\boldsymbol{f}$ can also be computed by changing the sign of the imaginary part of $\boldsymbol{f}$. The quantum states can thus be obtained for a pair of $i$th and $j$th samples, which are denoted by the green and the blue arrows, respectively. Re and Im in the figure denote the real and imaginary parts of a complex number, respectively. For the block diagram of complex multiplier (CM), see panel b of Supplementary Figure 5.

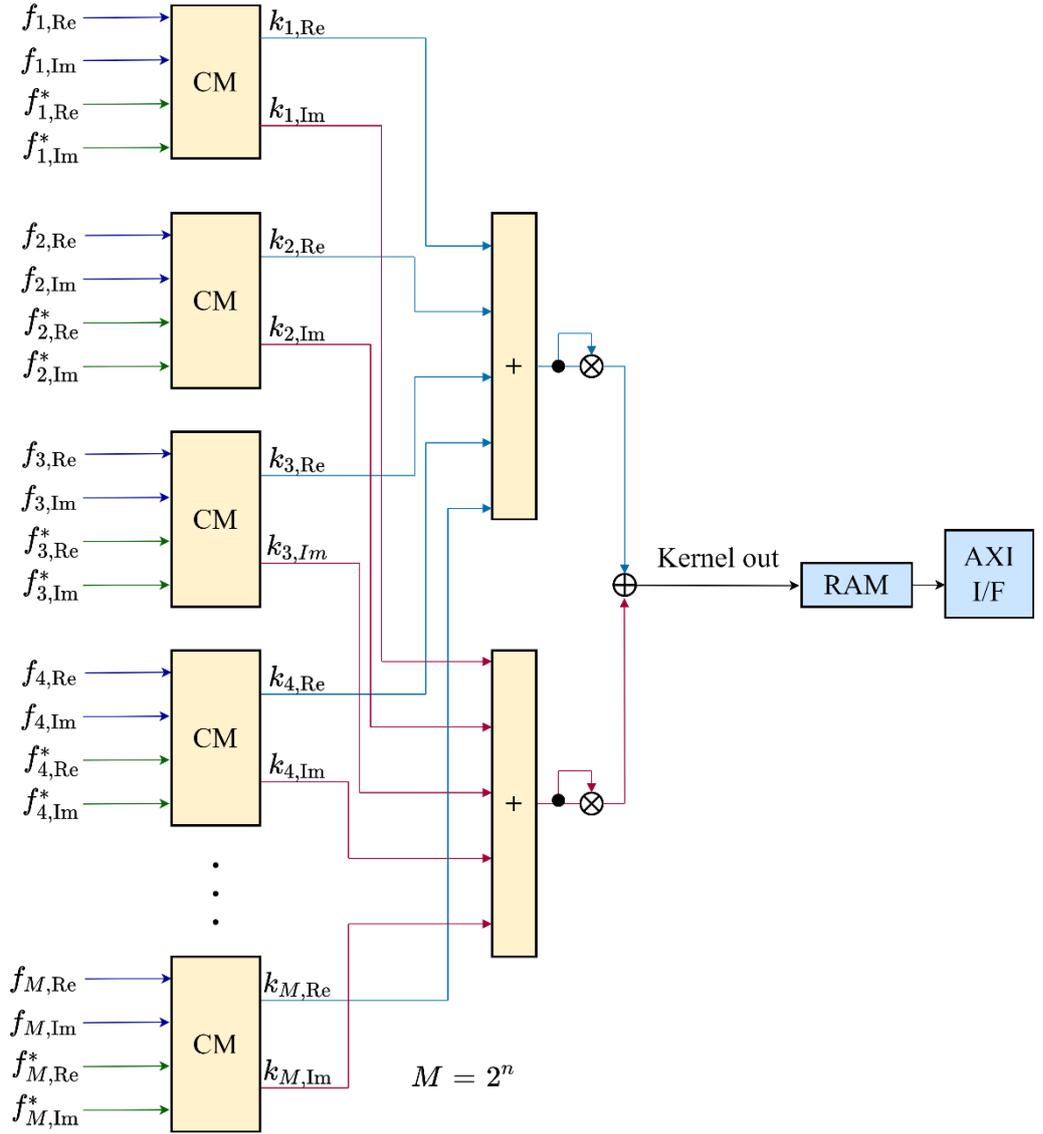

Supplementary Figure 7: Block diagram for the module that computes the square of the norm of the inner product. The inner product can be obtained using complex multipliers: $k_l := f_l^*(x^{(i)})f_l(x^{(j)}) \in \mathbb{C}$ $(l = 1, \cdots, M)$ for a pair of $i$th and $j$th samples, which are denoted by the green and the blue arrows, respectively. Then the quantum kernel entry $K_{ij}$ is given by the sum of $(\sum_l \text{Re}[k_l])^2$ and $(\sum_l \text{Im}[k_l])^2$, which are indicated by the light-blue and the red arrows, respectively. The data are stored in RAM and sent back to the host. For the block diagram of complex multiplier (CM), see part b of Supplementary Figure 5.

Supplementary Table 1: Hardware utilizations for the quantum kernel implementation

| Chip | XCVU9P (AWS F1 instance) | XCVU9P (AWS F1 instance) |
|---|---|---|
| Number of qubits | 2 | 6 |
| Train size | 1024 | 1024 |
| LUT | 175077/1180984 (15%) | 237681/1180984 (20%) |
| LUTRAM | 17335/591440 (3%) | 57360/591440 (10%) |
| FF | 250576/2364480 (11%) | 327521/2364480 (14%) |
| BRAM | 365.5/2160 (17%) | 515/2160 (24%) |
| URAM | 43/960 (4%) | 43/960 (4%) |
| DSP | 61/6840 (1%) | 1154/6840 (17%) |
| Clock frequency (MHz) | 250 | 250 |

Abbreviations: LUT, look-up table; RAM, random access memory; FF, flip flop; BRAM, block RAM; URAM, ultraRAM; DSP, digital signal processor.

Supplementary Note 3. Grid search over the hyperparameters for the classical and quantum kernels

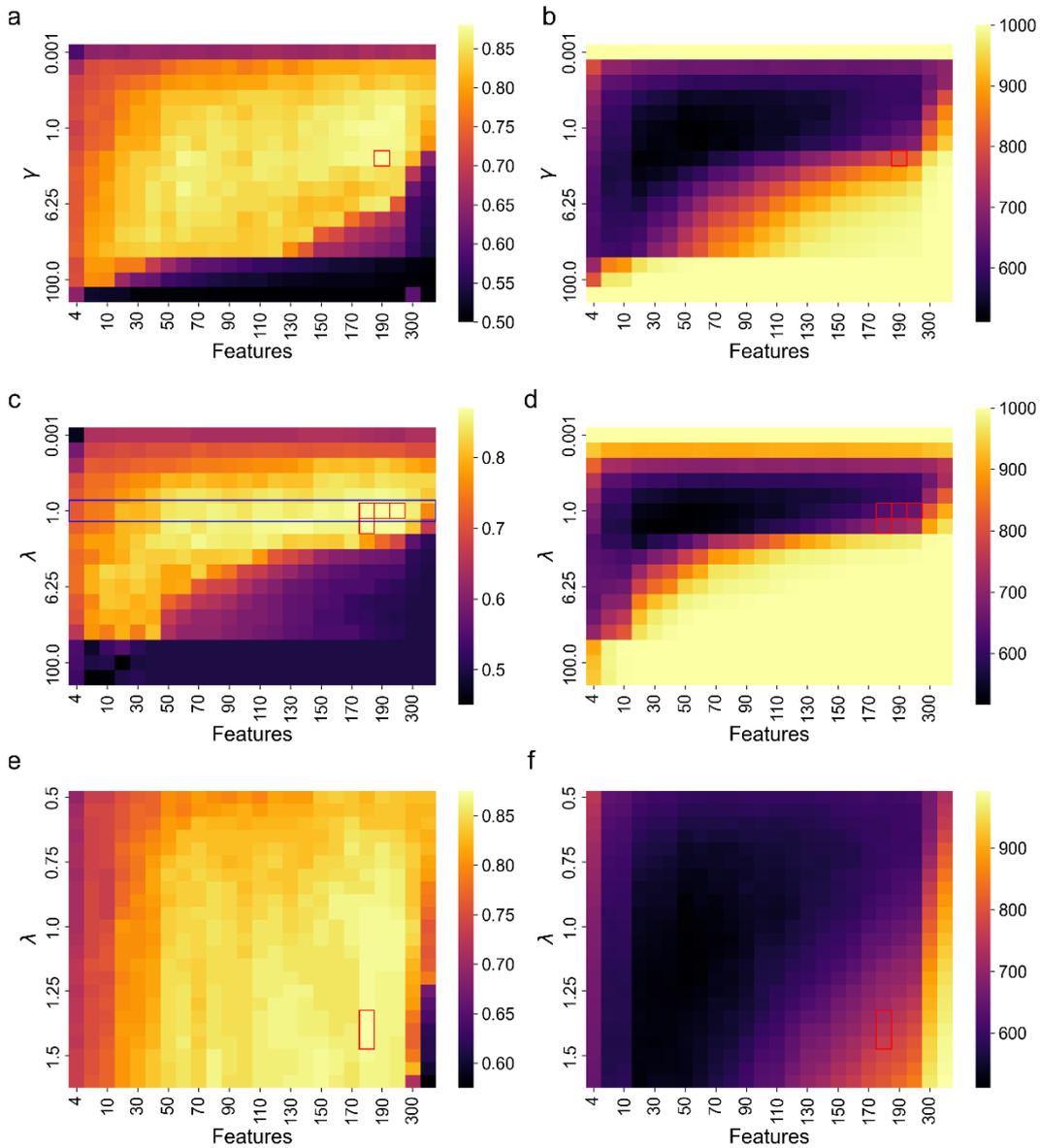

Supplementary Figure 8: Grid search over the hyperparameters for the classical and the quantum SVM on binary classification as a function of features. Fashion-MNIST dataset was used. The hyperparameters giving the optimal test accuracy are indicated by the open red square. (a) Grid search over the hyperparameter $\gamma$ for the Gaussian kernel: (a) test accuracy and (b) the number of the corresponding support vectors. Grid search over the scaling parameter $\lambda$ for the quantum kernel. (c) test accuracy and (d) the number of the corresponding support vectors. Note that, in our quantum kernel (panel c), the case of $\lambda = 1$ typically gave the near-optimal performance (indicated by the open blue rectangle), implying that our quantum kernel gave a reasonable performance without introducing any hyperparameter. Nonetheless, to further optimize the value for $\lambda$, we narrowed the range for $\lambda$ and performed another grid search over the scaling parameter: (e) test accuracy and (f) the number of the corresponding support vectors. A slightly better was obtained.

## Supplementary Note 4. Details of numerical simulations

We used the scikit-learn library [S2] for support vector machines. In order to compare our FPGA implantation with a quantum computing simulator, we used Qiskit, an open source software development kit [S3] (ver. 0.31.0) and Qiskit Aer (ver. 0.9.1) to perform quantum computing simulations in order to generate the quantum kernel in Fig. 2d in the main text.

## Supplementary References

[S1] Volder, J. E. The CORDIC trigonometric computing technique. *IRE Trans. Electron. Comput.* **3**, 330–334 (1959).

[S2] Pedregosa, F. *et al.* Scikit-learn: Machine learning in Python. *J. Mach. Learn. Res.* **12**, 2825–2830 (2011).

[S3] Aleksandrowicz, G. *et al.* Qiskit: An open-source framework for quantum computing. https://github.com/qiskit (Accessed September 13, 2022).